\documentclass[%
 aip,
 amsmath,amssymb,
 reprint,%
]{revtex4-1}

\usepackage{graphicx}
\usepackage{dcolumn}
\usepackage{bm}

\usepackage{amsmath}
\usepackage[utf8]{inputenc}
\usepackage[T1]{fontenc}
\usepackage{mathptmx}
\usepackage{etoolbox}
\usepackage{lineno,hyperref}
\usepackage{mathtools}
\usepackage{caption}
\usepackage{subcaption}

\makeatletter
\def\@email#1#2{%
 \endgroup
 \patchcmd{\titleblock@produce}
  {\frontmatter@RRAPformat}
  {\frontmatter@RRAPformat{\produce@RRAP{*#1\href{mailto:#2}{#2}}}\frontmatter@RRAPformat}
  {}{}
}%
\makeatother
\begin{document}

\preprint{AIP/123-QED}

\title{Numerical implications of including drifts in SOLPS-ITER simulations of EAST}
\author{D. Boeyaert}
\email{boeyaert@wisc.edu}
 \affiliation{University of Wisconsin-Madison, Department of Nuclear Engineering and Engineering Physics, 1500 Engineering Drive, Madison, WI 53706, United States of America}
  \affiliation{KU Leuven, Department of Mechanical Engineering, Celestijnenlaan 300, 3001 Leuven, Belgium}
  \affiliation{Forschungszentrum Jülich GmbH, Institut für Energie- und Klimaforschung – Plasmaphysik, 52425 Jülich, Germany}
\author{S. Carli}%
 \affiliation{KU Leuven, Department of Mechanical Engineering, Celestijnenlaan 300, 3001 Leuven, Belgium}

\author{W. Dekeyser}
 \affiliation{KU Leuven, Department of Mechanical Engineering, Celestijnenlaan 300, 3001 Leuven, Belgium}

\author{S. Wiesen}
\affiliation{Forschungszentrum Jülich GmbH, Institut für Energie- und Klimaforschung – Plasmaphysik, 52425 Jülich, Germany}

\author{M. Baelmans}
 \affiliation{KU Leuven, Department of Mechanical Engineering, Celestijnenlaan 300, 3001 Leuven, Belgium}

\date{\today}

\begin{abstract}
The inclusion of drifts in plasma edge codes like SOLPS-ITER is required to match simulation data with experimental profiles. However, this remains numerically challenging. In this paper, the effect of some numerical factors on the final plasma solution is investigated. This study is performed on three EAST simulations in upper single null configuration: an attached purely deuterium case, an attached case with limited Ne-seeding and a detached Ne-seeded case. The effects of the anomalous conductivity and anomalous thermo-electric coefficient on the plasma potential are investigated. Next, the effect of the employed grids is shown. In order to investigate these effects, accurate drift simulations are needed. Therefore, the employed time step and used numerical parameters are discussed for the three studied simulations. For all presented simulations, it is verified that the restriction of the grid to the first flux surface tangent to the main chamber wall has only a small effect on the divertor solution. This means that the main power dissipation takes place inside the simulated domain and only a small fraction of the power is leaving the B2.5 grid trough the grid boundary closest to the first wall. Finally the effect of drifts on the asymmetry between the inner and outer target for EAST simulations is demonstrated.
\end{abstract}

\maketitle

%

\section{Introduction}

Modeling of the plasma edge is widely used to understand which physical mechanisms can help reducing the power and particle exhaust towards the divertor targets in tokamaks \cite{wischmeier2015high}. For that, transport equations based on the ones of Braginskii are often solved. The SOLPS-ITER code package couples the finite volume plasma multi-fluid code B2.5 and the Monte Carlo kinetic neutral code EIRENE \cite{wiesen2015new,bonnin2016presentation}. B2.5 solves the fluid transport equations for the plasma, and EIRENE calculates the source terms appearing in these equations resulting from interactions with neutrals.

Drift effects cause up-down pressure asymmetries and radial ion flows \cite{chankin2013role}. This gives, in H-mode experiments, an asymmetry between the inner and outer divertor targets as observed experimentally in ref. \cite{reimold2014experimental}. In refs. \cite{dekeyser2017solps,wensing2020experimental, paradela2020impact, casali2022impurity} it is shown for different tokamaks that drifts are required to obtain this asymmetry in SOLPS-ITER profiles: higher density and particle fluxes are observed at the inner target -- the so-called high field side high density region --, while higher temperatures and heat fluxes are observed at the outer target. To analyze further in detail the Ne-seeding experiments at EAST from which a first analysis was performed in ref. \cite{boeyaert2021towards}, SOLPS-ITER simulations including drifts are required. 

This paper focuses on the numerical implications of turning on drifts in SOLPS-ITER while physical interpretation of the results and comparison with experimental data will be dealt with elsewhere. As indicated in ref. \cite{rozhansky2009new} the numerical schemes in SOLPS-ITER become unstable for drift simulations when an H-mode transport barrier is imposed in the simulation. 
Therefore, several artificial numerical parameters are introduced in the set of equations solved in SOLPS-ITER.

The effects of these artificial numerical transport coefficients is verified. Additionally, the effects of space and time discretization, and of the coupling to a Monte Carlo code are analyzed. It has been shown that these factors influence the outcome of SOLPS-ITER simulations from the numerical side, and in that way might influence drifts-enabled simulations \cite{bonnin2020solps,ghoos2018grid,kaveeva2018speed}.


The analysis is performed for the deuterium and Ne-seeded experiments from ref. \cite{boeyaert2021towards}. Furthermore, the numerical implications of including Ne in an attached simulation are checked.

The performed analysis limits to simulations for the EAST tokamak. For the first time, the effect of these numerical parameters is studied in detail for a SOLPS-ITER simulation. The findings are not necessary valid for simulations of other devices, but as shown later on, some findings agree with published results elsewhere. Therefore, the presented outcome can help to obtain drift-enabled simulations for other devices.

The examined EAST discharges are in a disconnected double null configuration with main upper divertor. However, as the separation between the two separatrices is $\sim 2 \, \mathrm{cm}$, they are considered as discharges in an upper single null configuration. If the upper single null topology is used during the grid generation for the SOLPS-ITER grids, this implies that the grids only extend $2 \, \mathrm{cm}$ into the SOL outside the separatrix. As the power decay length for the examined experiments is estimated to be maximum $5 \, \mathrm{mm}$ \cite{boeyaert2021towards}, a grid which goes $2 \, \mathrm{cm}$ into the SOL covers 4 times the power decay length which should be sufficient.

The paper is organized as follows: in section 2, the way the electric currents are calculated in SOLPS-ITER is briefly discussed. In section 3, the effects of the choice of some numerical input parameters on the profiles of SOLPS-ITER simulations including drifts for attached purely deuterium, attached Ne-seeded simulations, and for the detached Ne-seeded case are presented. In section 4, the impact of drifts on the distribution of the power deposition is examined, before coming to a summary of the main results.

\section{Model description}

In ref. \cite{rozhansky2009new} the basic equations solved in the SOLPS-ITER code are given. They consist of the continuity equation, the parallel momentum conservation equation, and the ion and electron energy conservation equations. This set of equations can be solved while neglecting drift and electric current effects. To include drift effects on the particle flows in the simulations, the charge continuity equation has to be additionally solved. 

The charge continuity equation ($\mathbf{\nabla} \cdot \mathbf{j} = 0$) is formulated as follows in SOLPS-ITER (using the toroidal symmetry of tokamaks):

\begin{equation}
	\frac{1}{\sqrt{g}}\frac{\partial}{\partial x} \left(\frac{\sqrt{g}}{h_x} \tilde{j_x} \right) + \frac{1}{\sqrt{g}} \frac{\partial}{\partial y} \left( \frac{\sqrt{g}}{h_y} \tilde{j_y} \right) = 0,
	\label{eq:current_continuity}
\end{equation}

in which $x$ is the poloidal direction and $y$ the radial direction normal to the magnetic flux surfaces. $h_x$ is the cell width in the poloidal direction, $h_y$ the cell width in the radial direction. $g$ indicates the surface area between a grid cell and its neighbor. 
In this equation $\mathbf{\tilde{j}}$ is the plasma current vector and it is split in the following way:


\begin{equation}
	\mathbf{\tilde{j}} = \mathbf{j}_{AN} + \mathbf{\tilde{j}}_{dia} + \mathbf{j}_{in} + \mathbf{\tilde{j}}_{vis||} + \mathbf{\tilde{j}}_{visq} + \mathbf{j}_{vis\perp} + \mathbf{j}_{s} + \mathbf{j}_{||}.
	\label{eq:current_components}
\end{equation}

This shows that the plasma current consists of an (artificial) anomalous ($\mathbf{\tilde{j}}_{AN}$), and a neoclassical current. The later can be further split up in the diamagnetic current ($\mathbf{\tilde{j}}_{dia}$), current caused by inertia and gyroviscosity ($\mathbf{j}_{in}$), viscosity contributions ($\mathbf{\tilde{j}}_{vis||} + \mathbf{\tilde{j}}_{visq} + \mathbf{j}_{vis\perp}$), neutral-ion friction contributions ($\mathbf{j}_{s}$), and a parallel current originating from the electron parallel momentum balance ($\mathbf{j}_{||}$) \cite{rozhansky2001simulation}. The tilde on some of the terms indicates that divergence free parts of these currents are eliminated analytically \cite{rozhansky2003potentials}. $\mathbf{\tilde{j}}_{dia}$, $\mathbf{\tilde{j}}_{vis||}$, and $\mathbf{\tilde{j}}_{visq}$ represent physically a vertical guiding center drift caused by the curvature of the magnetic field $\mathbf{B}$ and $\nabla B$.  Important here is to state the meaning of the artificial anomalous current. Often "anomalous" plasma parameters are referred to as a result of turbulence. In the presented paper, however, "(artificial) anomalous" is used to denote artificial transport quantities that are introduced to enhance the numerical stability and convergence of the code. This choice is made to use the same terminology as employed in the SOLPS-ITER manual \cite{bonnin2020solps}.

In modeled H-mode discharges with plasma edge codes like SOLPS-ITER, typically a transport barrier is imposed for the perpendicular diffusion coefficient \cite{senichenkov2019mechanisms}. This results in small values for the imposed perpendicular diffusion for a large fraction of the simulation domain. As a result, solving equation \ref{eq:current_continuity} is difficult from a numerical viewpoint. Therefore, the artificial anomalous current $\mathbf{\tilde{j}}_{AN}$ is introduced which is dependent on the gradient of the potential.
\begin{equation}
	\begin{split}
		j_{AN,x} = - \sigma_{AN} \frac{1}{h_x} \frac{\partial \phi}{\partial x} \\
		j_{AN,y} = - \sigma_{AN} \frac{1}{h_y} \frac{\partial \phi}{\partial y}
	\end{split}
\end{equation}
This artificial anomalous current is not physical, but is only present to ensure smooth convergence. As a result, its contribution should be large enough to ensure convergence but at the same time small to avoid a non-physical simulation result. Therefore, an appropriate choice of the artificial anomalous conductivity $\sigma_{AN}$ should be made.  This means that $\sigma_{AN} << \sigma_{NEO} \approx \frac{\mu_{i1}}{(BR)^2}$ with $\mu_{i1}$ the parallel viscosity \cite{kaveeva2018speed}. In SOLPS-ITER, $\sigma_{AN}$ is determined by $c_{\sigma_{\alpha,0}}$ with the following formula \cite{bonnin2020solps}:
\begin{equation}
	{\sigma }_{AN}=c_{{\sigma }_{\alpha },0}\cdot e\ n_e
	\label{eq:an_cond_1}
\end{equation}
in which $n_e$ is the electron density at the innermost grid surface at the outer midplane. Further possible dependencies of ${\sigma }_{AN}$ are described in ref. \cite{bonnin2020solps} but are not considered in the scope of the presented work.

The value of the neoclassical conductivity depends on the collision frequency regime in which the tokamak is operating. 
Therefore, it is often easier to ensure that $\mathbf{j}_{AN} << \mathbf{j}_{NEO}$ at each location. 

As indicated earlier, the last right-hand term of equation \ref{eq:current_components} ($\mathbf{j}_{||}$) is originating from the electron parallel momentum balance. If no impurities are present in the plasma, this current is given by the following equation \cite{bonnin2020solps}:
\begin{equation}
		j_{||,x} = \sigma_{||} \left(\frac{1}{en_eh_x}\frac{\partial n_e T_e}{\partial x}  -  \frac{1}{h_x}\frac{\partial \phi}{\partial x}\right) - \alpha_{ex} \frac{1}{h_x}\frac{\partial T_e}{\partial x} \\
	\label{eq:j_||,x}
\end{equation}
in which $\sigma_{||}$ is the Spitzer conductivity and $\phi$ the plasma potential. The first part of the right hand side compensates for the diamagnetic current, and forms the Pfirsch-Schlüter current. The second term on the right is the thermo-electric current which arises due to temperature gradients. In this expression, $\alpha_{ex}$ is the thermo-electric coefficient. This coefficient consists of a classical part, and an artificial anomalous contribution. As for the artificial anomalous current, the artificial anomalous thermo-electric coefficient is:
\begin{equation}
	\alpha_{AN} = c_{\alpha,0} \cdot e\ n_e
\end{equation}

Similar to the artificial anomalous current, the artificial anomalous thermo-electric coefficient is only present for convergence purposes. Therefore, its contribution should be small in comparison with the classical thermo-electric coefficient: $\alpha_{AN} << \alpha_{CL}$. 


To close the charge continuity equation (equation \ref{eq:current_continuity}), boundary conditions (BCs) should be imposed. At the far scrape-off layer, it is common to impose a zero gradient for $\phi$ \cite{sytova2020modeling,wensing2021drift}. At the core boundary, the radial electric current is put equal to the diamagnetic current \cite{rozhansky2009new,sytova2020modeling}, and at the private flux boundaries of the grid, the artificial anomalous current is put to zero. At the divertor targets, sheath BCs are imposed. As this results in a non-linear BC, a modified implementation of sheath BCs for drift simulations has been developed \cite{bonnin2020solps}. As mentioned in ref. \cite{reimold2014experimental}, the imposed sheath conditions limit the numerical stability of the code.  

Once the plasma potential and the magnetic field are known, the drift velocities resulting from the $\mathbf{E}$ x $\mathbf{B}$ drift and the diamagnetic drift velocities resulting from the $\mathbf{\nabla B}$ drift can be calculated:
\begin{equation}
	V^{E \, \mathrm{x} \, B}_x = - \frac{B_z}{B^2}  \frac{1}{h_y}\frac{\partial \phi}{\partial y}, \,
	V^{E \, \mathrm{x} \, B}_y = \frac{B_z}{B^2}  \frac{1}{h_x}\frac{\partial \phi}{\partial x}
	\label{eq:drift_velocity_ExB}
\end{equation}
\begin{equation}
	V^{dia}_x =  \frac{T_i B_z}{e}  \frac{1}{h_y}\frac{\partial}{\partial y} \frac{1}{B^2}, \,
	V^{dia}_y =  - \frac{T_i B_z}{e} \frac{1}{h_x}\frac{\partial}{\partial x} \frac{1}{B^2}.
	\label{eq:drift_velocity_dia}
\end{equation}

\begin{figure}[h]
	\centering
	\includegraphics[width=8cm]{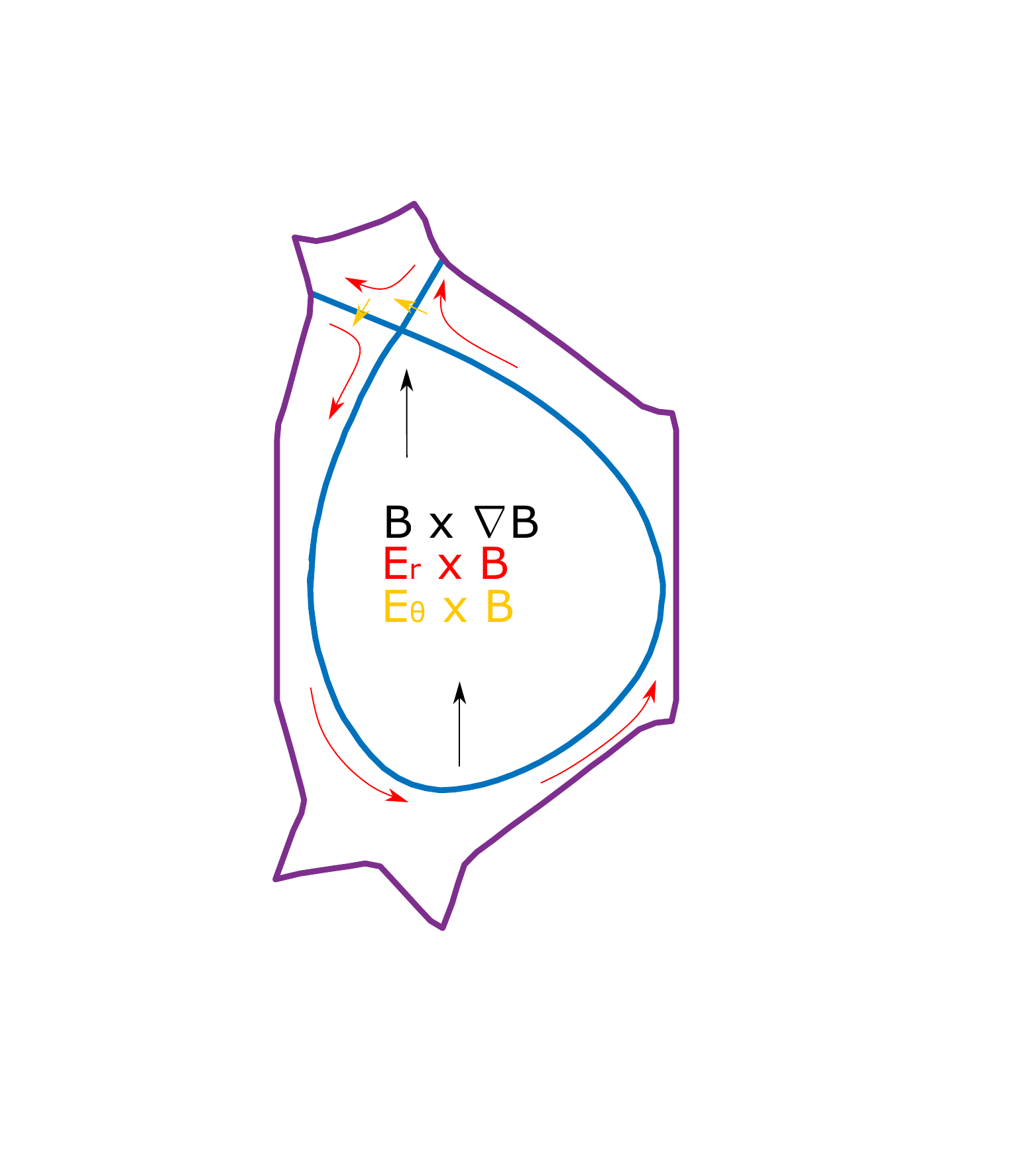}
	\caption[Drift flows]{The direction of the (ion) drift flows in the studied EAST discharge. Three different flows are indicated: the one caused by the magnetic field and its derivative (in black), the one caused by the radial electric field in combination with the magnetic field (in red) and the one caused by the poloidal electric field and the magnetic field (yellow). The influence of the radial electric field is larger than the one of the poloidal electric field.}
	\label{fig:USN_drifts}  
\end{figure}

As the drift velocity will influence the overall velocity of the plasma particles, it also influences the other four transport equations 
(particle balance, parallel momentum balance, and energy equations for electrons and ions) as demonstrated in ref. \cite{rozhansky2009new}. Therefore, their computation has a large influence on the examined plasma quantities, and is further investigated through this paper.  For the studied EAST case, the directions of the appearing drift flows caused by the electric and magnetic field are indicated in figure \ref{fig:USN_drifts}. 

Starting from the SOLPS-ITER setups of ref. \cite{boeyaert2022numerical} for EAST, drifts are turned on. For the presented simulations the same convergence criteria as in ref. \cite{boeyaert2022numerical} are used. In the entire paper the 3.0.7 master version of the SOLPS-ITER code is employed.

\section{Including drifts in SOLPS-ITER EAST simulations}

Equation \ref{eq:j_||,x} can be rewritten as:
\begin{equation}
	\frac{1}{h_x}\frac{\partial \phi}{\partial x} = - \eta_{||} j_{||,x} - \eta_{||}\alpha_{ex} \frac{1}{h_x}\frac{\partial T_e}{\partial x} + \frac{1}{en_eh_x}\frac{\partial n_e T_e}{\partial x} 
	\label{eq:potential}
\end{equation}
in which $\eta_{||} = \frac{1}{\sigma_{||}}$. Under attached conditions, the Spitzer conductivity will be high \cite{pitcher1997experimental} meaning that the first right-hand term of equation \ref{eq:potential} is small. The basic two-point model in attachment shows that $2 n_t T_t = n_u T_u$ in which 'u' refers to the upstream conditions and 't' to the target conditions \cite{stangeby2000plasma}. This makes that also the last right-hand term is small, resulting in:
\begin{equation}
	\frac{\partial \phi}{\partial x} \sim \frac{\partial T_e}{\partial x}
\end{equation}
This shows that the local potential is determined by the integration of the temperature along a flux tube, and that the potential at a specific location is directly linked to the electron temperature at that location.
Under attached conditions, the maximum temperature at a divertor target is larger than in detachment and the temperature gradients in the perpendicular direction are significant. As a consequence, large perpendicular gradients in the plasma potential will be present. This makes the charge continuity equation more unstable from a numerical viewpoint.

\begin{figure}[h]
	\centering
	\begin{subfigure}{8cm}
		\includegraphics[width=8cm]{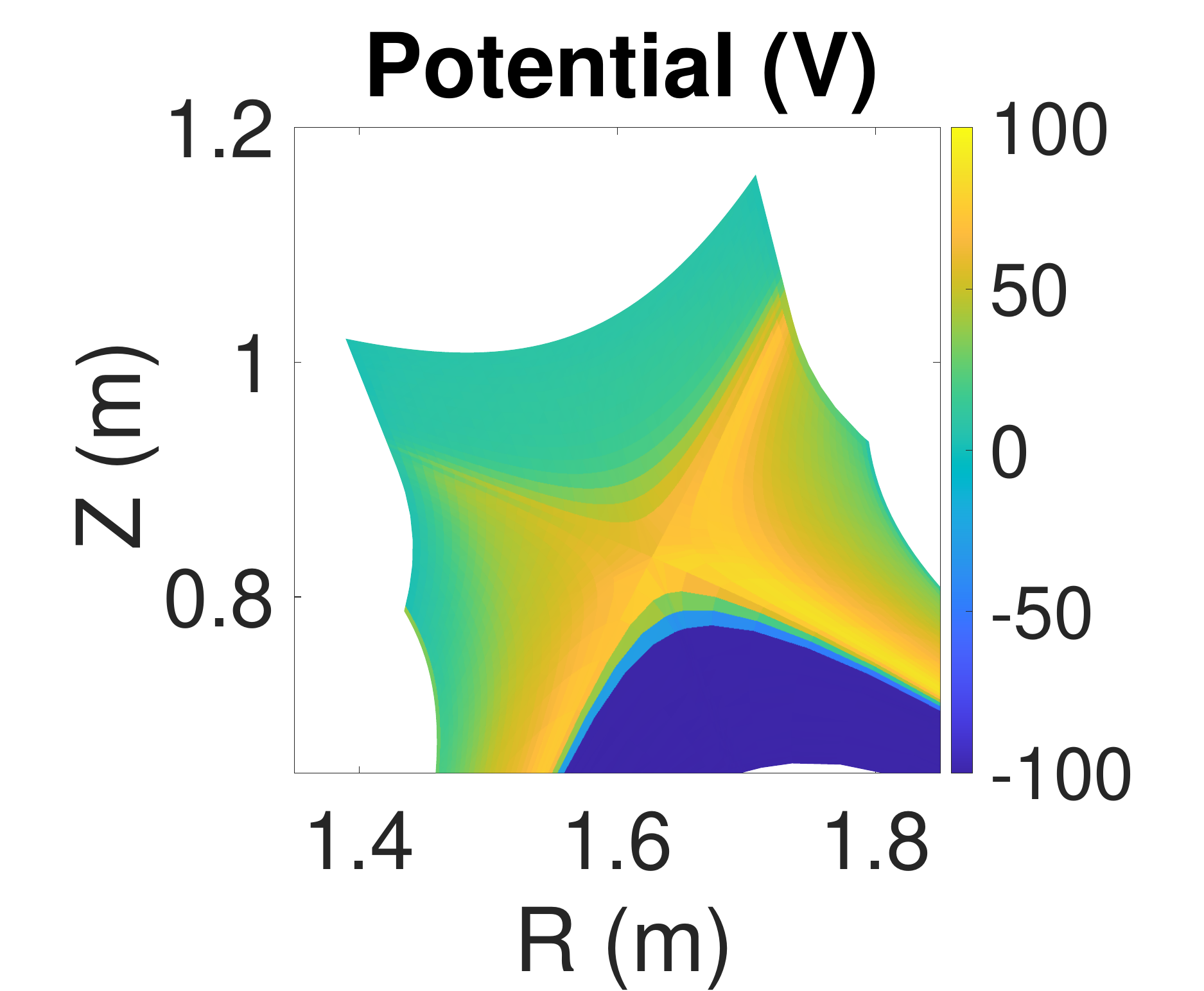}
		\caption{}
		\label{subfig:plasma_potential_D2}
	\end{subfigure}
	\begin{subfigure}{8cm}
		\includegraphics[width=8cm]{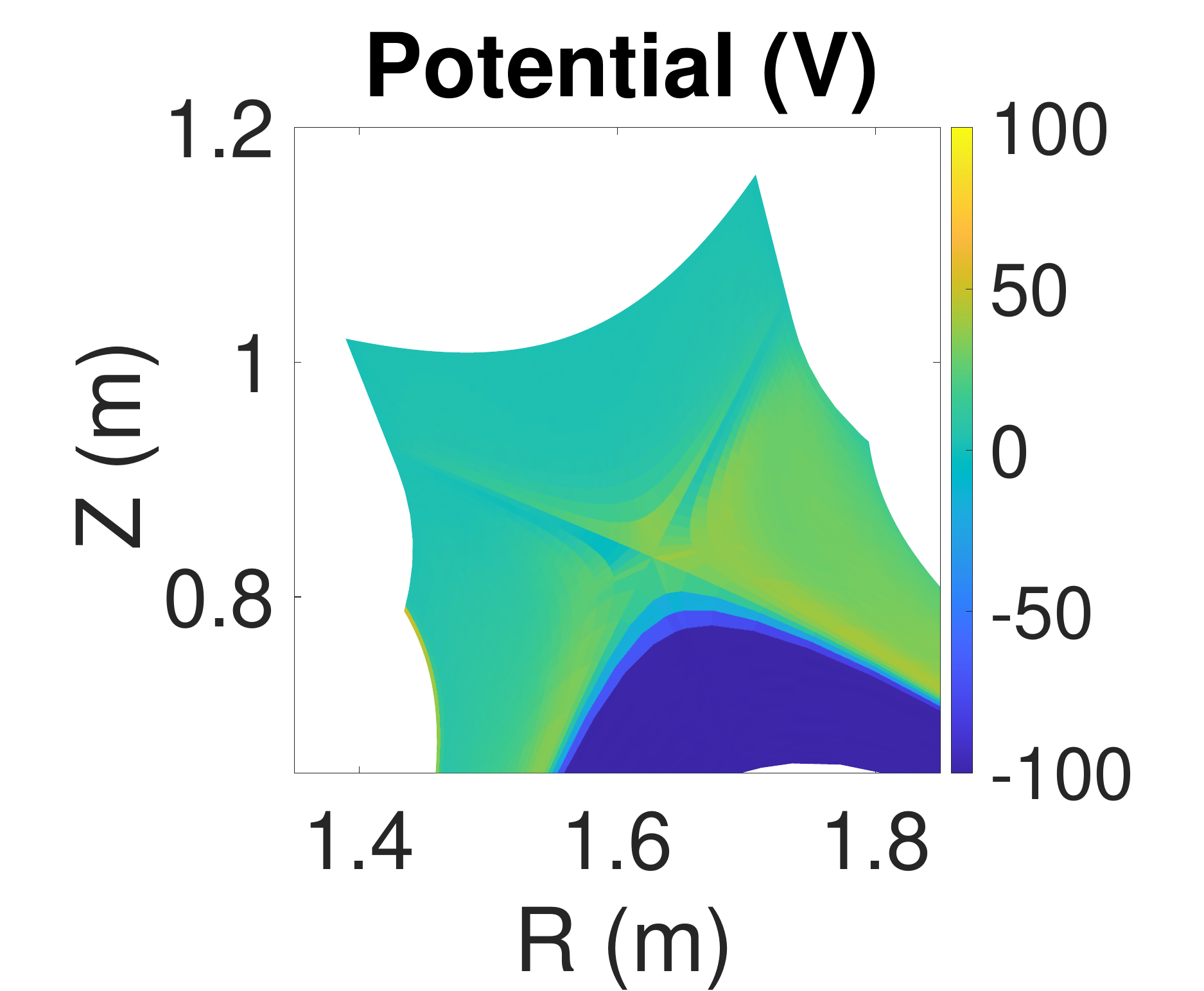}
		\caption{}
		\label{subfig:plasma_potential_Ne}
	\end{subfigure}
	\caption{The plasma potential for a simulation in the high-recycling regime (a) and the one for a detached simulation (b)}
	\label{fig:plasma_potential}
\end{figure}

This is demonstrated in figure \ref{fig:plasma_potential}. The variations in plasma potential in the divertor region for a simulation in attached conditions are larger than the variations for a detached simulation. Therefore, it is expected that drifts are easier to include in a detached simulation than in an attached one.

Switching on all terms of the charge continuity equation at once does not lead to a converged solution. Whereas the viscous terms could be turned on fully straight away, the other drift terms (their strengths) have to be gradually added, step by step. First only $10 \%$ of drift terms, afterwards $50 \%$, $80 \%$ and finally $100 \%$. Only when the case with $10 \%$ drifts is converged, the output is used to start the case with $50 \%$ activated drifts, and so on.

The methods described in ref. \cite{kaveeva2018speed} are applied to the performed EAST simulations to enhance the inclusion of drifts: working with an intermediate solution by using a higher $\sigma^{(AN)}$ (and decreasing it again for the last simulation steps achieving final convergence), separating the time derivatives for the energy and density between inside and outside the separatrix, and including an artificial particle source to balance pumping and puffing more easily. Especially separating the time derivatives is necessary to obtain convergence for the presented simulations. More details about this method can be found in ref. \cite{kaveeva2018speed}.

In this section, first the inclusion of drifts in attached purely deuterium simulations is discussed. Afterwards Ne-seeding is added but the plasma is still attached. This changes some of the investigated numerics. Finally, enough Ne-seeding is included to bring the simulation in detachment. As indicated before, including drifts in detached cases is easier than in attached ones.


\subsection{Including drifts in a purely deuterium attached simulation}

It is verified how the process of including drifts can be done in the most efficient way for a purely deuterium SOLPS-ITER simulation of EAST. The investigated simulation is the reference case from refs. \cite{boeyaert2021towards, boeyaert2022numerical}. Equations \ref{eq:drift_velocity_ExB} and \ref{eq:drift_velocity_dia} show that the computation of drift velocities is influenced by two factors: the $\mathbf{\phi}$ and $\mathbf{B}$ fields at one hand, and the calculation of their derivatives on the other. First, the influence of the imposed artificial anomalous conductivity and thermo-electric coefficient on the $\mathbf{\phi}$ field is investigated. As these artificial anomalous terms affect the solved set of equations, they modify the investigated physical problem. Afterwards, the influence of the involved numerics is investigated by the effect of the grid on the computation of $\mathbf{\phi}$. 
In a next step, the influence of the involved number of EIRENE particles and the averaging over simulation steps is discussed, before finishing with an analysis of the used time step and numerical switches in SOLPS-ITER. For all presented purely deuterium simulations, an input power of $2.05 \, \mathrm{MW}$ is imposed at the core boundary of the grid.

\subsubsection{Appropriate choice of artificial anomalous conductivity and thermo-electric coefficient}


A smooth profile of the plasma potential $\phi$ is required as discontinuities can hamper gradient calculations in equation \ref{eq:drift_velocity_ExB}.  $\phi$ is determined by the set of solved equations. Therefore, the influence of the choice of the artificial anomalous parameters - $\sigma_{AN}$ and $\alpha_{AN}$ - on the potential is investigated. 

\begin{figure}[h]
	\centering
	\includegraphics[width=7cm]{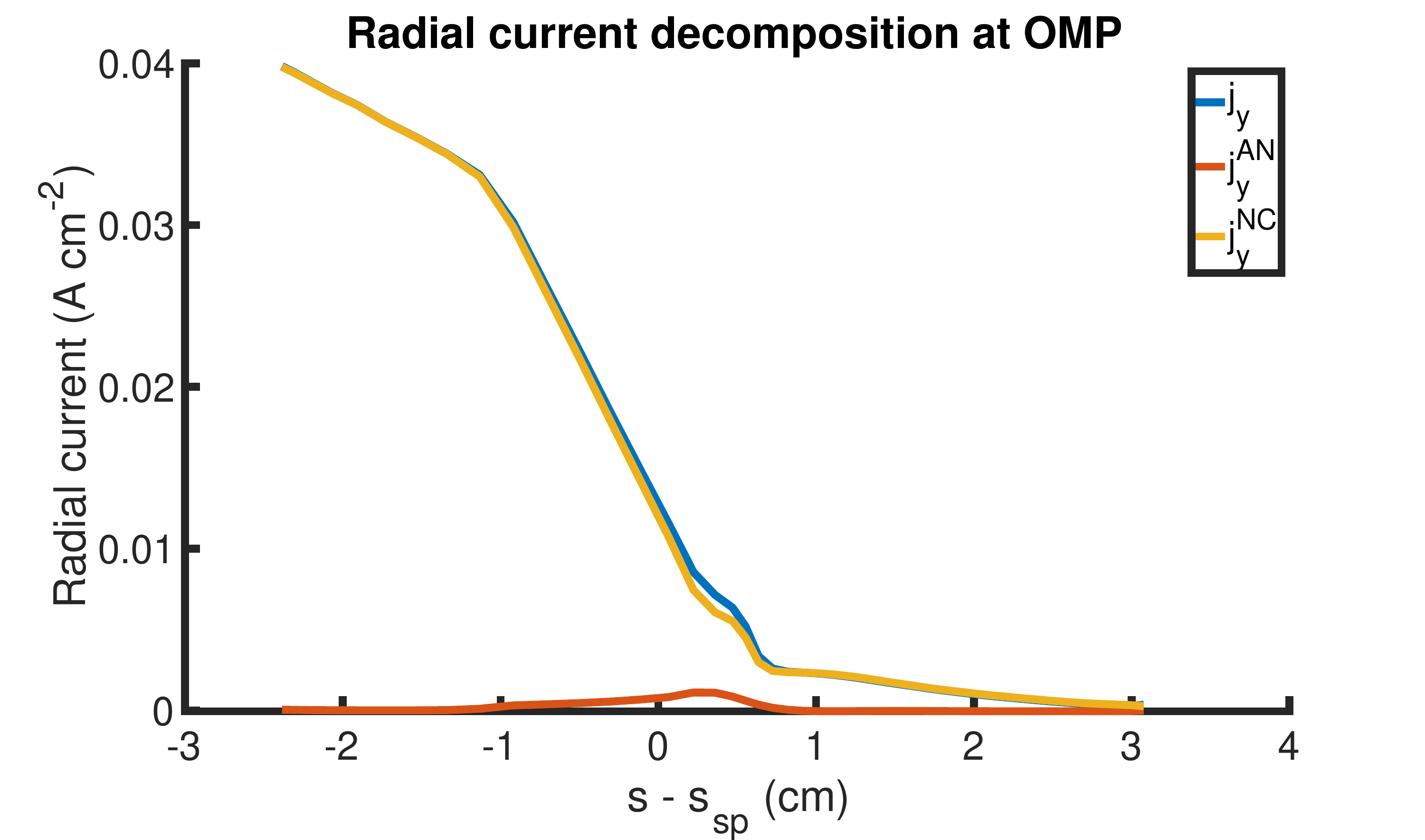}
	\caption{Artificial anomalous current contribution at the OMP.}
	\label{fig:jan_jncl}
\end{figure}

The artificial anomalous parameters which were successfully used to include drifts in ASDEX Upgrade \cite{sytova2020modeling}, are employed in the presented work. This results in $c_{\sigma_{\alpha,0}} = 5.0 \cdot 10^{-5}$ and $c_{\alpha,0} = 1.0 \cdot 10^{-5}$. This choice is explained by the fact that ASDEX Upgrade has a similar size as EAST. In figure \ref{fig:jan_jncl} it is verified that $\sigma_{AN} << \sigma_{NEO}$ by examining that $\mathbf{\tilde{j}}_{AN} << \mathbf{\tilde{j}}_{NEO}$ for the simulations discussed above at the OMP. Similar results are obtained at the UOT.

First the sensitivity of $\sigma_{AN}$ on $\phi$  is investigated, afterwards the impact of $\alpha_{AN}$ is studied. 

While changing the values of $c_{\sigma_{\alpha,0}}$ the impact on the convergence becomes clear. Where a simulation in which $c_{\sigma_{\alpha,0}}$ is increased to $10.0 \cdot 10^{-5}$ converges within a couple of days, it was not possible to get a converged solution in case of $c_{\sigma_{\alpha,0}} \le 4.0 \cdot 10^{-5}$ in which the solution for the density and temperature profiles kept oscillating and did not became stable. 

\begin{figure}[h]
	\centering
	\includegraphics[width=7cm]{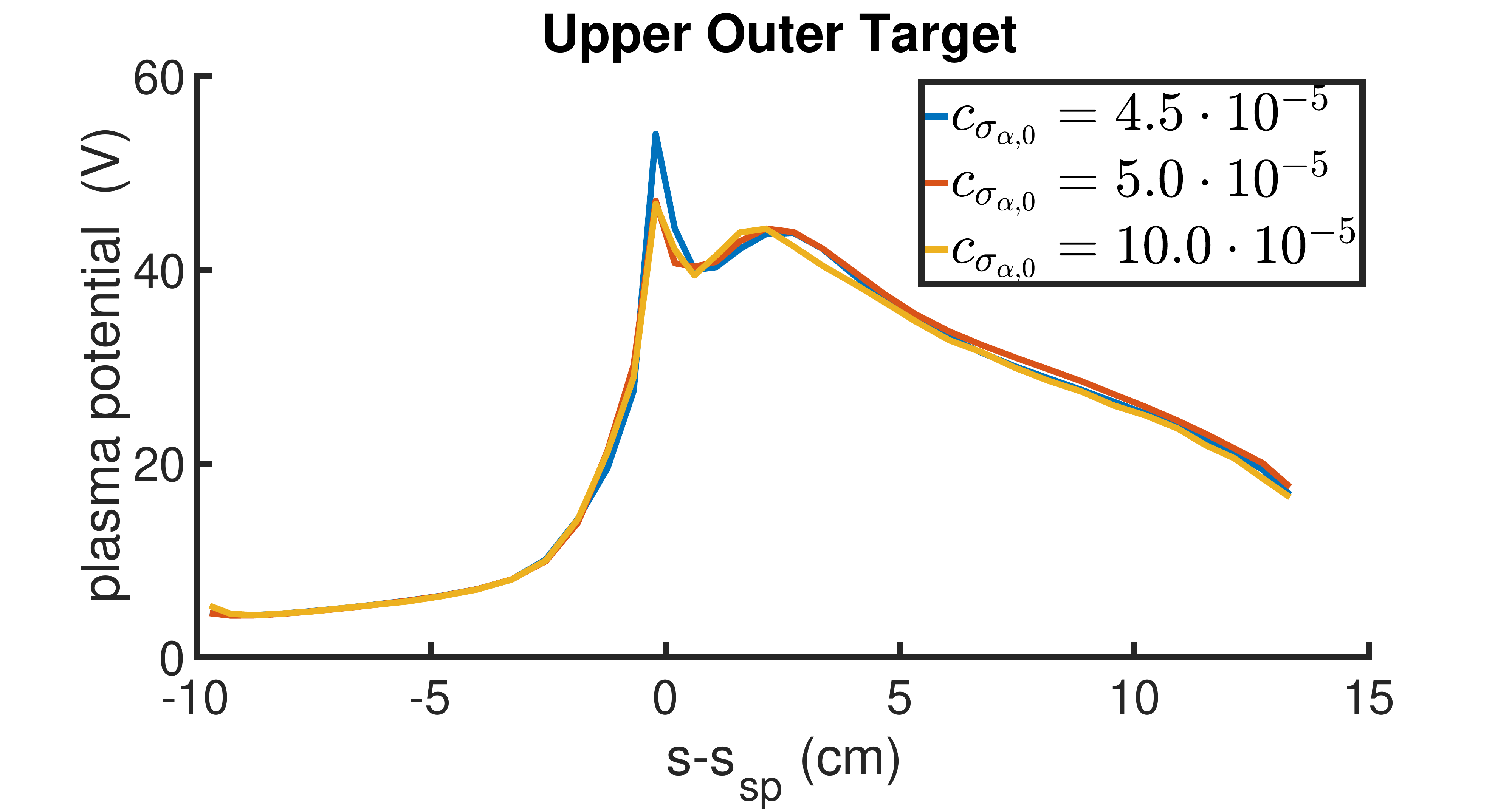}
	\caption{The influence of the choice of $c_{\sigma_{\alpha,0}}$ on plasma potential at the UOT.}
	\label{fig:UOT_potential_cfsig}
\end{figure}

In figure \ref{fig:UOT_potential_cfsig} the plasma potential is compared at the upper outer target (UOT) for different values of $c_{\sigma_{\alpha,0}}$.  While the convergence speed increases with increasing $\sigma_{\alpha,0}$, artificial smoothing of the potential target profile is observed. This indicates a smaller gradient of the target potential, and in that way an underestimation of the electric field. This means that we have to make a trade-off between accuracy and convergence speed.

\begin{figure}[h]
	\centering
	\includegraphics[width=7cm]{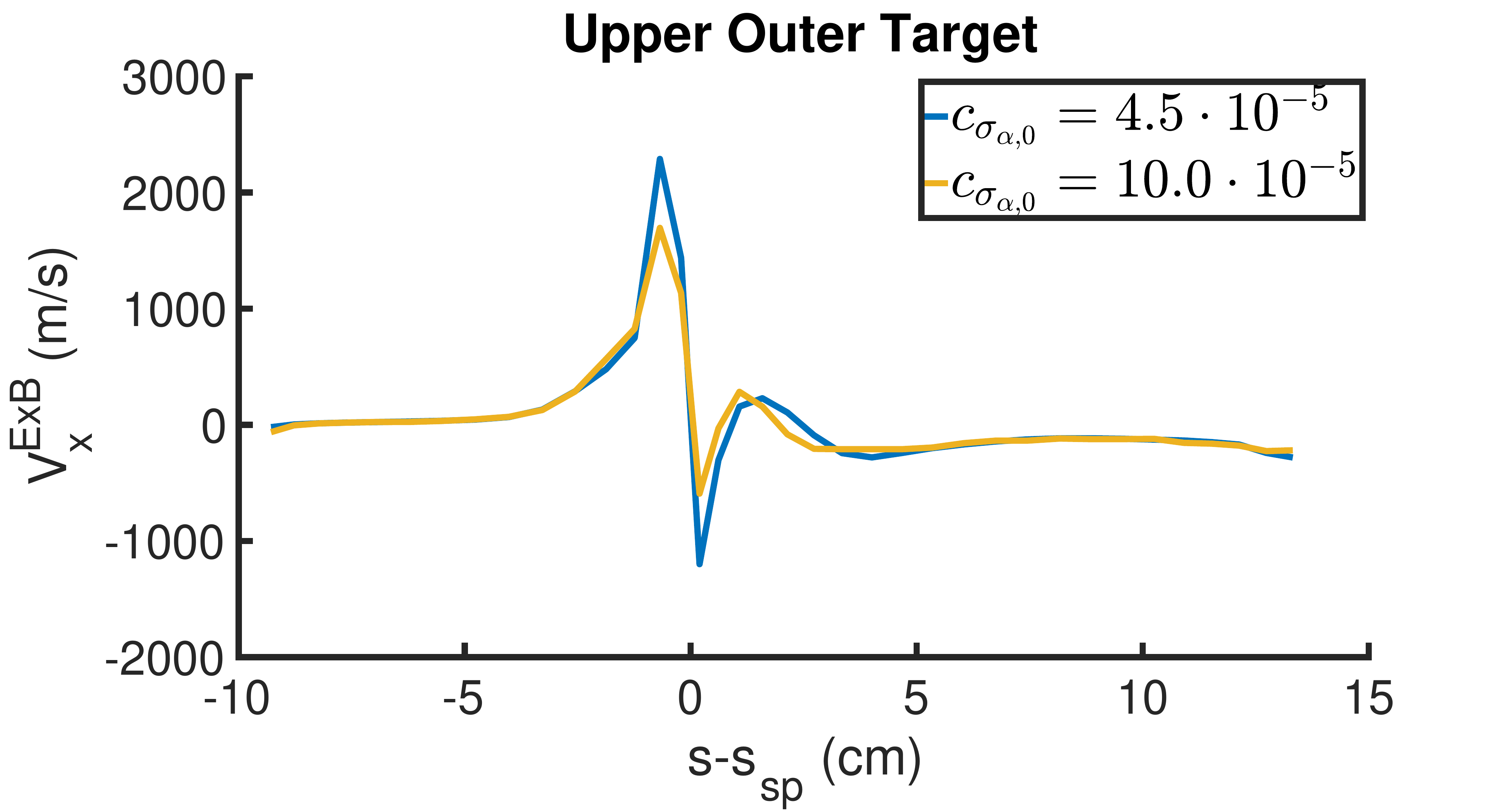}
	\caption{The influence of the choice of $c_{\sigma_{\alpha,0}}$ on the $V^{ExB}_x$ drift velocity at the UOT.}
	\label{fig:UOT_vExB_cfsig}
\end{figure}

The potential drop of figure \ref{fig:UOT_potential_cfsig} causes opposite poloidal ExB drift velocities across the separatrix at the UOT as shown in figure \ref{fig:UOT_vExB_cfsig}.

\begin{figure}[h]
	\centering
	\includegraphics[width=7cm]{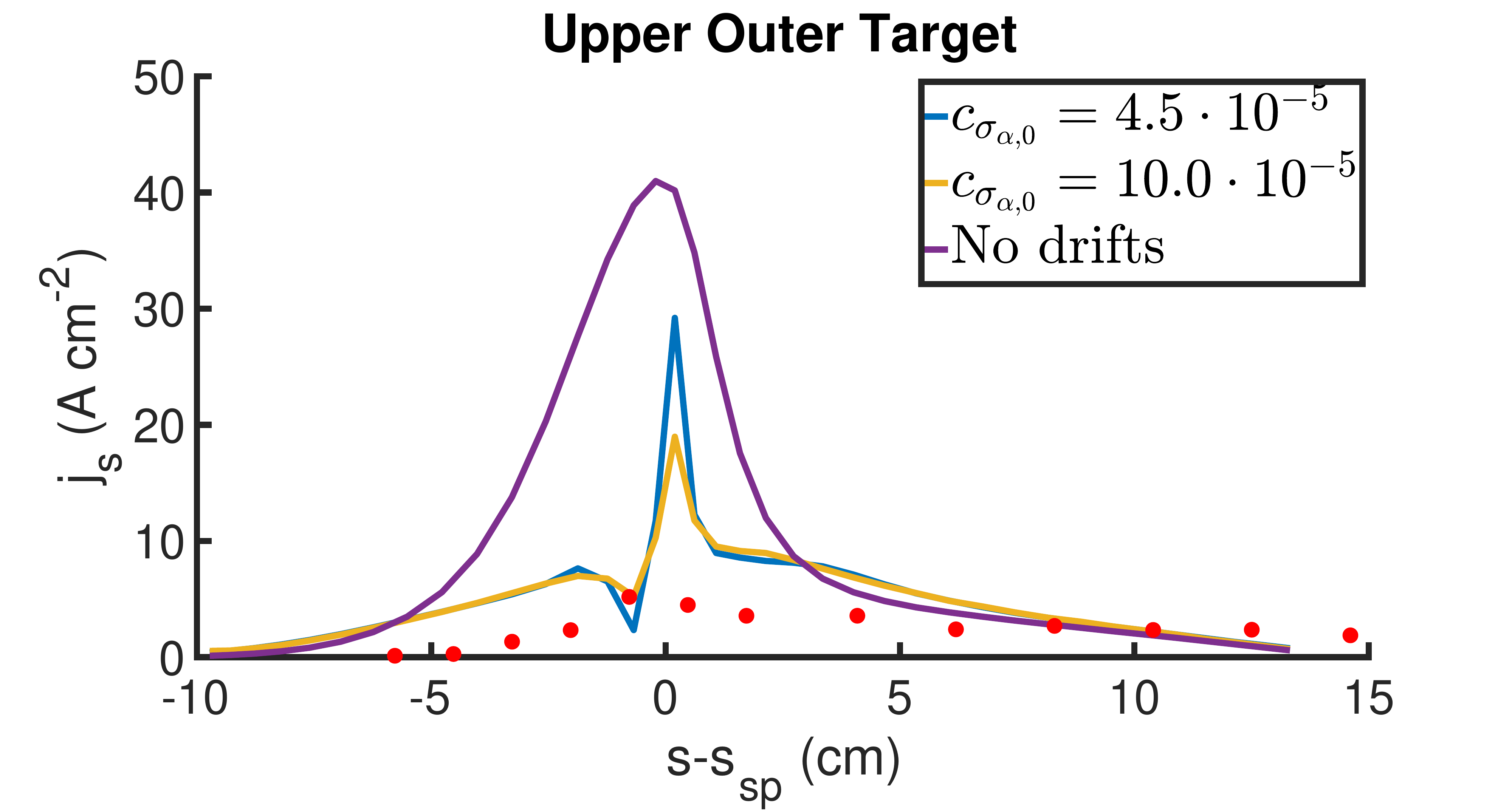}
	\caption{The influence of the choice of $c_{\sigma_{\alpha,0}}$ on the ion saturation current at the UOT. For reference the simulation result without drift effects is shown in purple, and the experimental data from the divertor Langmuir probe are given in red.}
	\label{fig:UOT_js_cfsig}
\end{figure}

As the ion saturation current can be compared with the divertor Langmuir probes at EAST, this is one of the key plasma quantities for comparison with experiment \cite{boeyaert2021towards}. The effect of the employed artificial anomalous conductivity on the ion saturation current at the UOT is shown in figure \ref{fig:UOT_js_cfsig}: for the simulation with $c_{\sigma_{\alpha,0}} = 10.0 \cdot 10^{-5}$ the peak value for the ion saturation current is $\sim 50 \cdot \%$ lower than for the simulation with $c_{\sigma_{\alpha,0}} = 4.5 \cdot 10^{-5}$, but figure \ref{fig:UOT_potential_cfsig} shows that this is because of an underestimation of the potential. 

In figure \ref{fig:UOT_js_cfsig} the simulation result without drift effects is shown in purple to show the influence of drifts on the $j_s$ profile. In red the measurements from the divertor Langmuir probes discussed in ref. \cite{boeyaert2021towards} are plotted. This indicates that the inclusion of drifts in the simulation improves the agreement between simulations and experiments which is the motivation for the effort presented in this paper to include drifts in SOLPS-ITER simulations.


In section \ref{subsubsec:spatial_discretization} the effect of small grid cells on the stability of the solution is demonstrated. A way to decrease the effect of such instabilities appearing with small grid cells, is changing $\sigma_{AN}$ depending on the grid cell width. In the 3.0.7 master version of SOLPS-ITER, the following scaling of $\sigma_{AN}$ with the grid cell width is available if $c_{\sigma_{\alpha,7}} \neq 0$:

\begin{equation}
	{\sigma }_{AN,new}=  c_{{\sigma }_{\alpha },7} \left(\frac{h_y}{h_{y,sep_{OMP}}} \right)^2  {\sigma}_{AN,old}
	\label{eq:an_cond_2}.
\end{equation}

\begin{figure}[h]
	\centering
	\includegraphics[width=7cm]{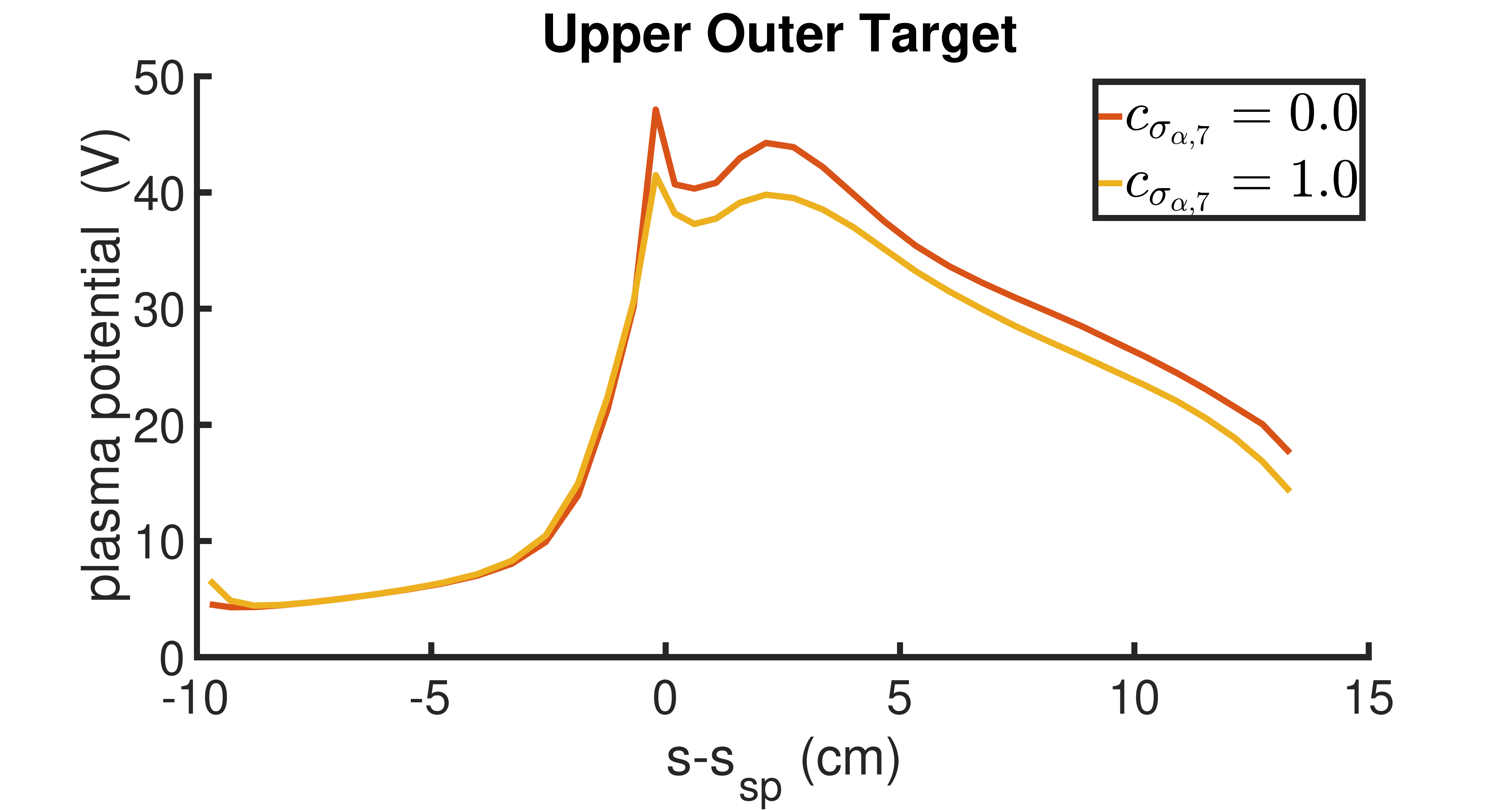}
	\caption{The influence of the choice of $c_{\sigma_{\alpha,7}}$ on plasma potential at the UOT.}
	\label{fig:UOT_potential_cfsig7}
\end{figure}

\begin{figure}[h]
	\centering
	\includegraphics[width=7cm]{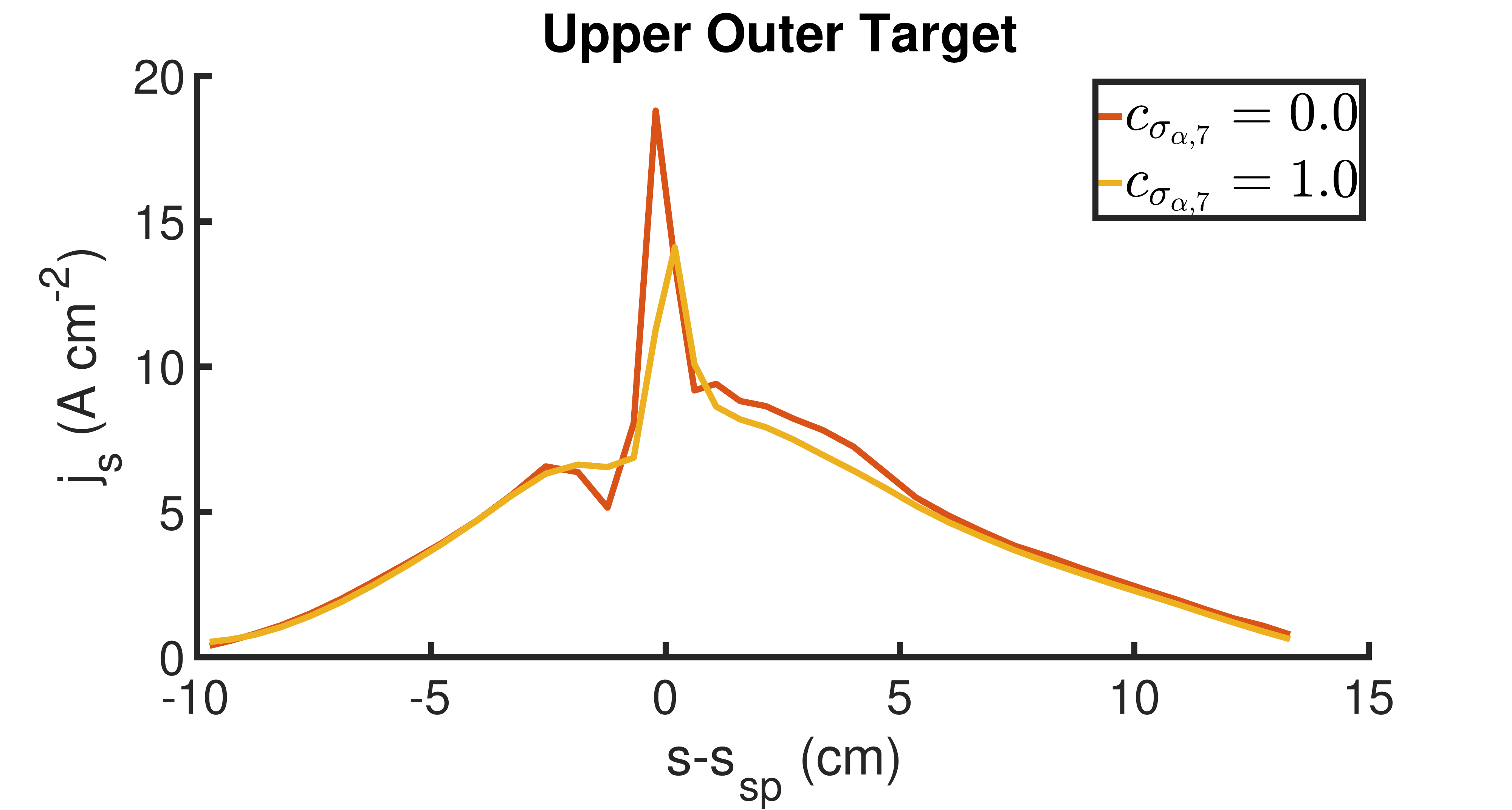}
	\caption{The influence of the choice of $c_{\sigma_{\alpha,7}}$ on the ion saturation current at the UOT.}
	\label{fig:UOT_js_cfsig7}
\end{figure}

As the grid cell width is smaller around the outer midplane (OMP) due to the flux expansion towards the targets, this means that the artificial anomalous conductivity is increased in the region around the targets. In this region, the effect will be similar to increasing $c_{\sigma_{\alpha,0}}$ as demonstrated before, but in the remaining part of the SOL, the effect will be smaller. Figure \ref{fig:UOT_potential_cfsig7} shows indeed that the potential drop is eliminated using $c_{\sigma_{\alpha,7}} = 1$.  Also in this case, it is easier to achieve convergence, but there is a loss of accuracy. The effect on the $j_s$ profile at the UOT is shown in figure \ref{fig:UOT_js_cfsig7}. It might be suggested this method is better than increasing $\sigma_{AN}$ everywhere as only the divertor region is affected.

\begin{figure}[h]
	\centering
	\includegraphics[width=7cm]{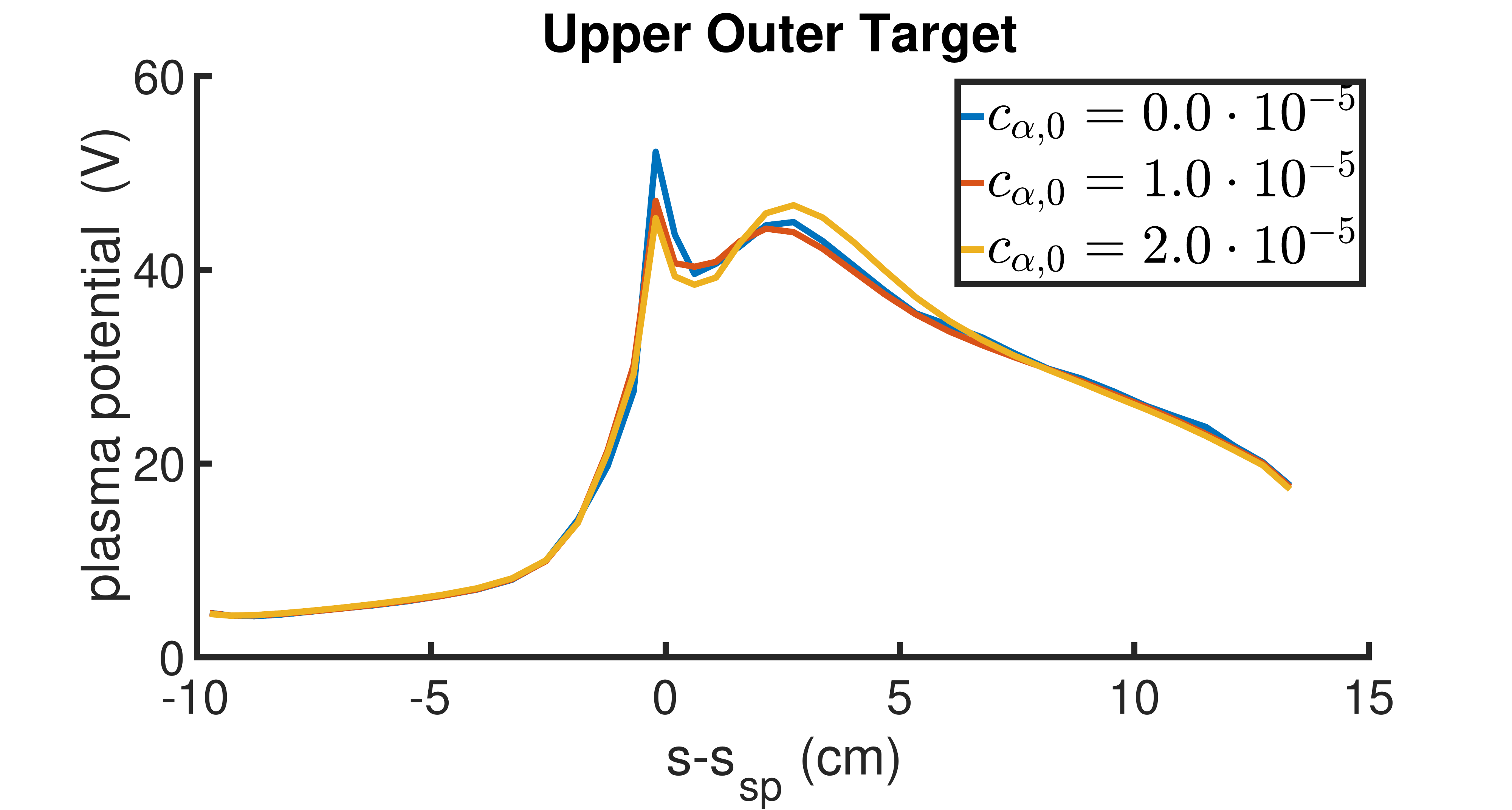}
	\caption{The influence of the choice of $c_{\alpha,0}$ on plasma potential at the UOT.}
	\label{fig:UOT_potential_cfalfa}
\end{figure}

\begin{figure}[h]
	\centering
	\includegraphics[width=7cm]{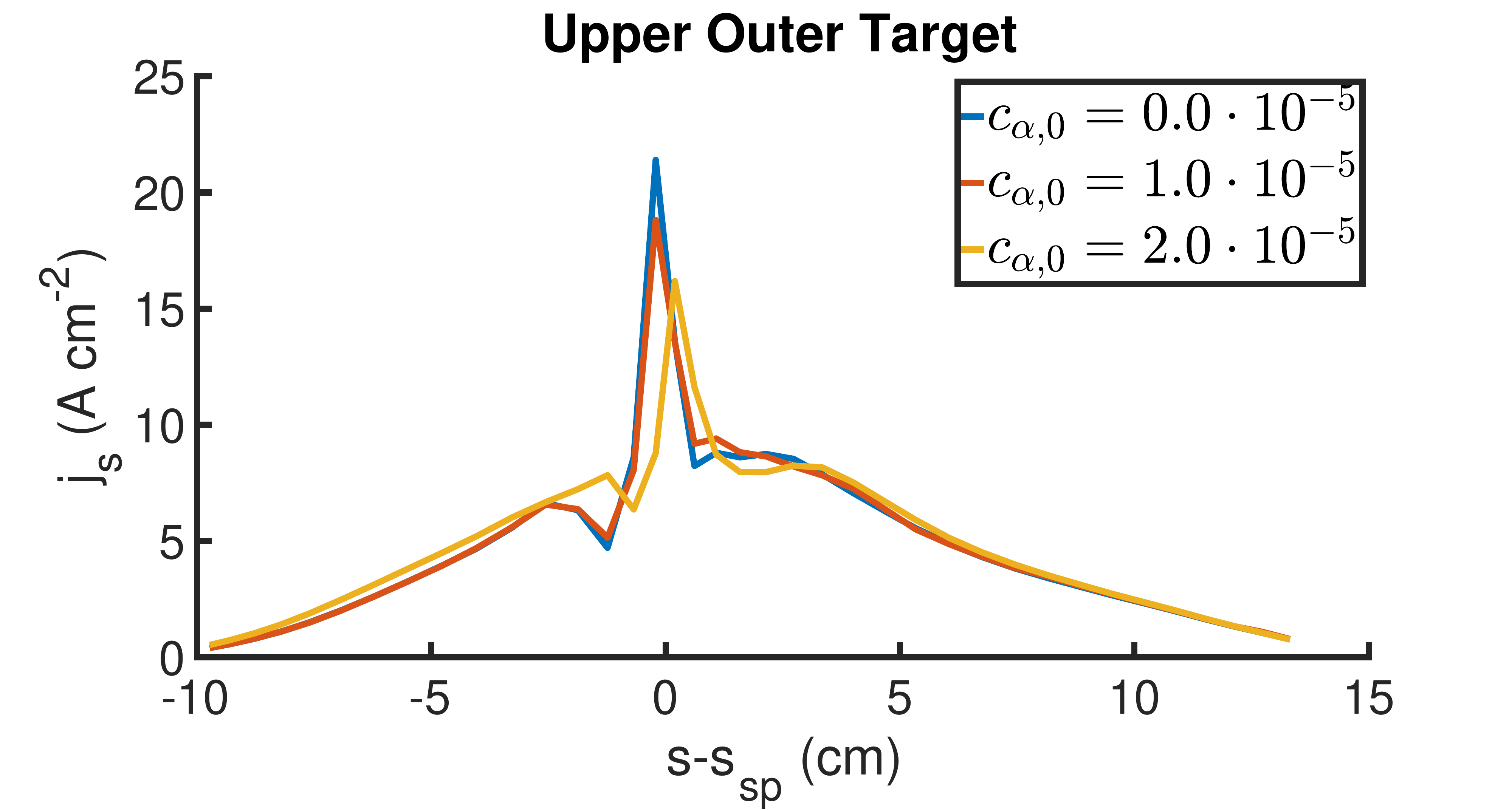}
	\caption{The influence of the choice of $c_{\alpha,0}$ on the ion saturation current at the UOT.}
	\label{fig:UOT_js_cfalfa}
\end{figure}

In a next step, the influence of $\alpha_{AN}$ is investigated. In ref. \cite{bonnin2020solps} it is stated that at least $\sigma_{AN}$ or $\alpha_{AN}$ cannot be zero if drift flows are included in a SOLPS-ITER calculation. As $\sigma_{AN}$ is fixed now to a finite value, the effects of putting $c_{\alpha,0} = 0.0 \cdot 10^{-5}$, $c_{\alpha,0} = 1.0 \cdot 10^{-5}$ and $c_{\alpha,0} = 2.0 \cdot 10^{-5}$ are examined. The resulting potentials are shown in figure \ref{fig:UOT_potential_cfalfa}. Also here, the simulations will converge faster for higher values of $c_{\alpha,0}$ but the potential profile is less well representing the drop around the separatrix. The effect on the ion saturation current at the UOT is shown in figure \ref{fig:UOT_js_cfalfa}.

\begin{figure}[h]
	\centering
	\includegraphics[width=7cm]{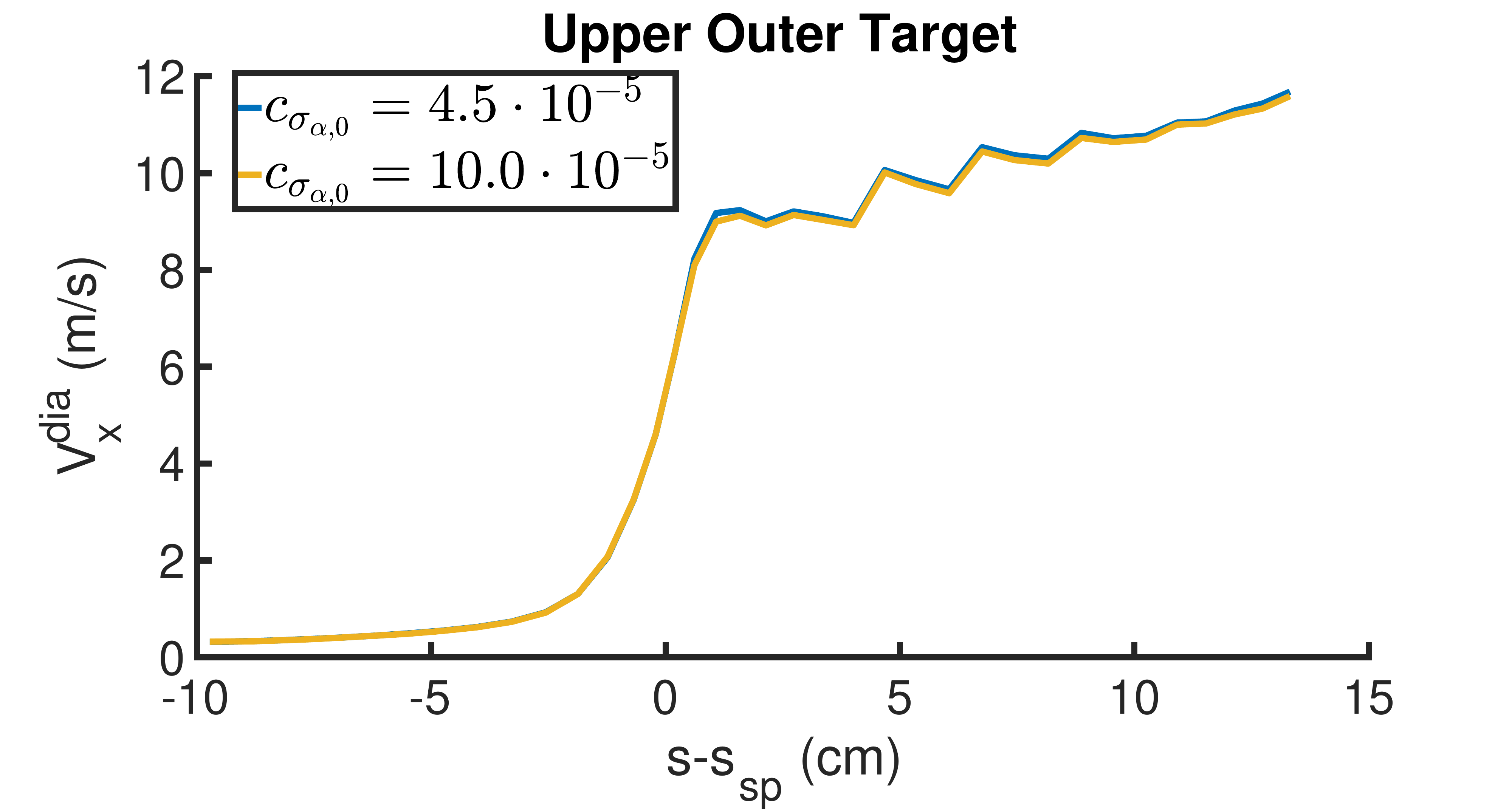}
	\caption{The influence of the choice of $c_{\sigma_{\alpha,0}}$ on the $V^{dia}_x$ drift velocity at the UOT.}
	\label{fig:UOT_vdia_cfsig}
\end{figure}

It should be remarked that the investigated artificial anomalous quantities have the largest influence on the $\mathbf{E}$ x $\mathbf{B}$ drifts. The diamagnetic drift velocities mainly depend on the magnetic field which is an input parameter for SOLPS-ITER and independent of the potential. Therefore, their values stay similar as indicated in figure \ref{fig:UOT_vdia_cfsig} for the diamagnetic drift profiles at the UOT for different values of $c_{\sigma_{\alpha,0}}$.

\subsubsection{Influence of the spatial discretization}
\label{subsubsec:spatial_discretization}


In the previous section it was shown that the artificial anomalous conductivity and thermo-electric coefficient can speed up convergence, but decrease the accuracy of the calculated solution.

In this section, the effect of the grid on the calculated solution is investigated.



\begin{figure}[h]
	\centering
	\includegraphics[width=7cm]{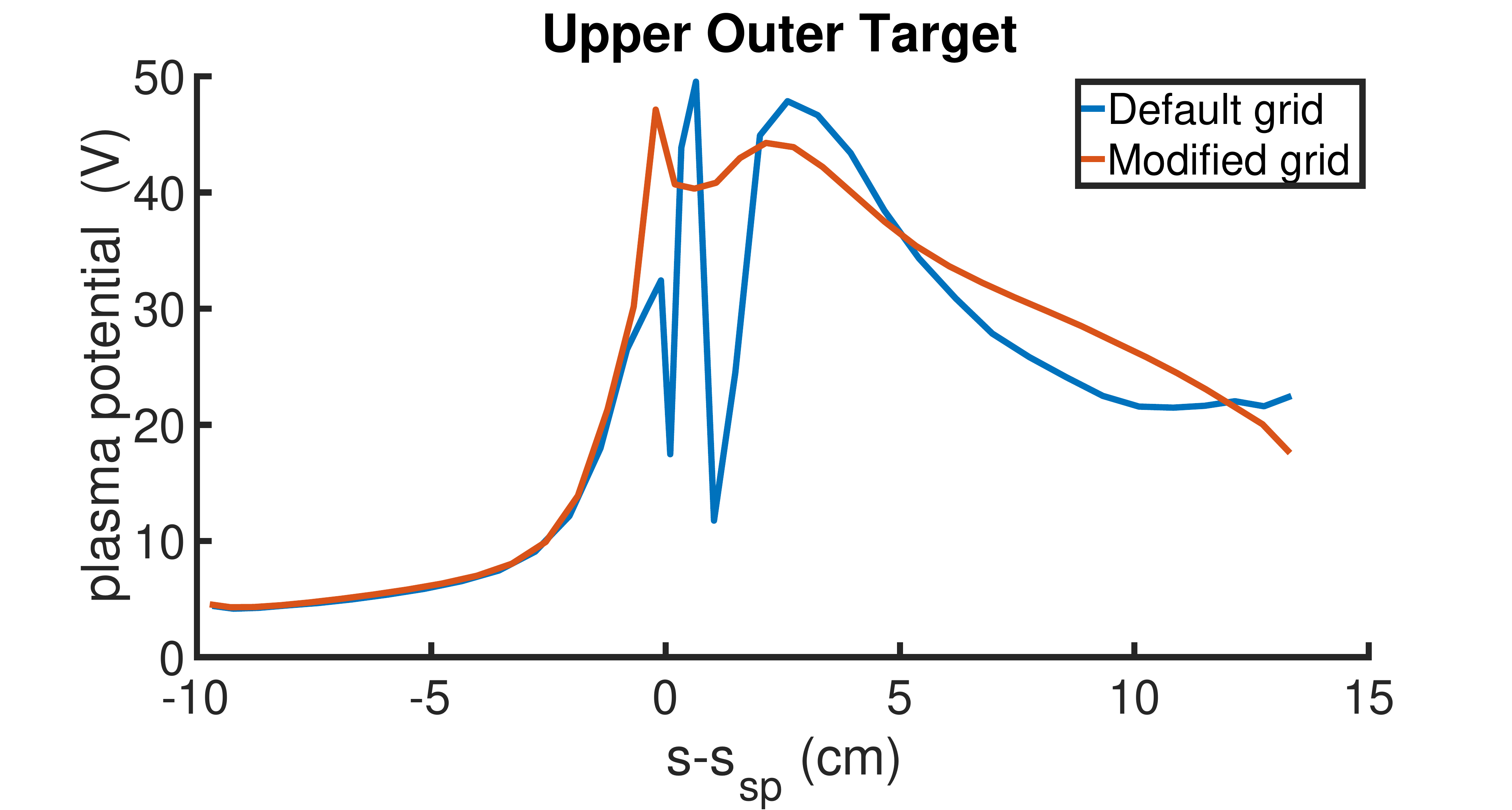}
	\caption{The influence of the grid on plasma potential at the UOT.}
	\label{fig:UOT_potential_grid}
\end{figure}
\begin{figure}[h]
	\centering
	\includegraphics[width=7cm]{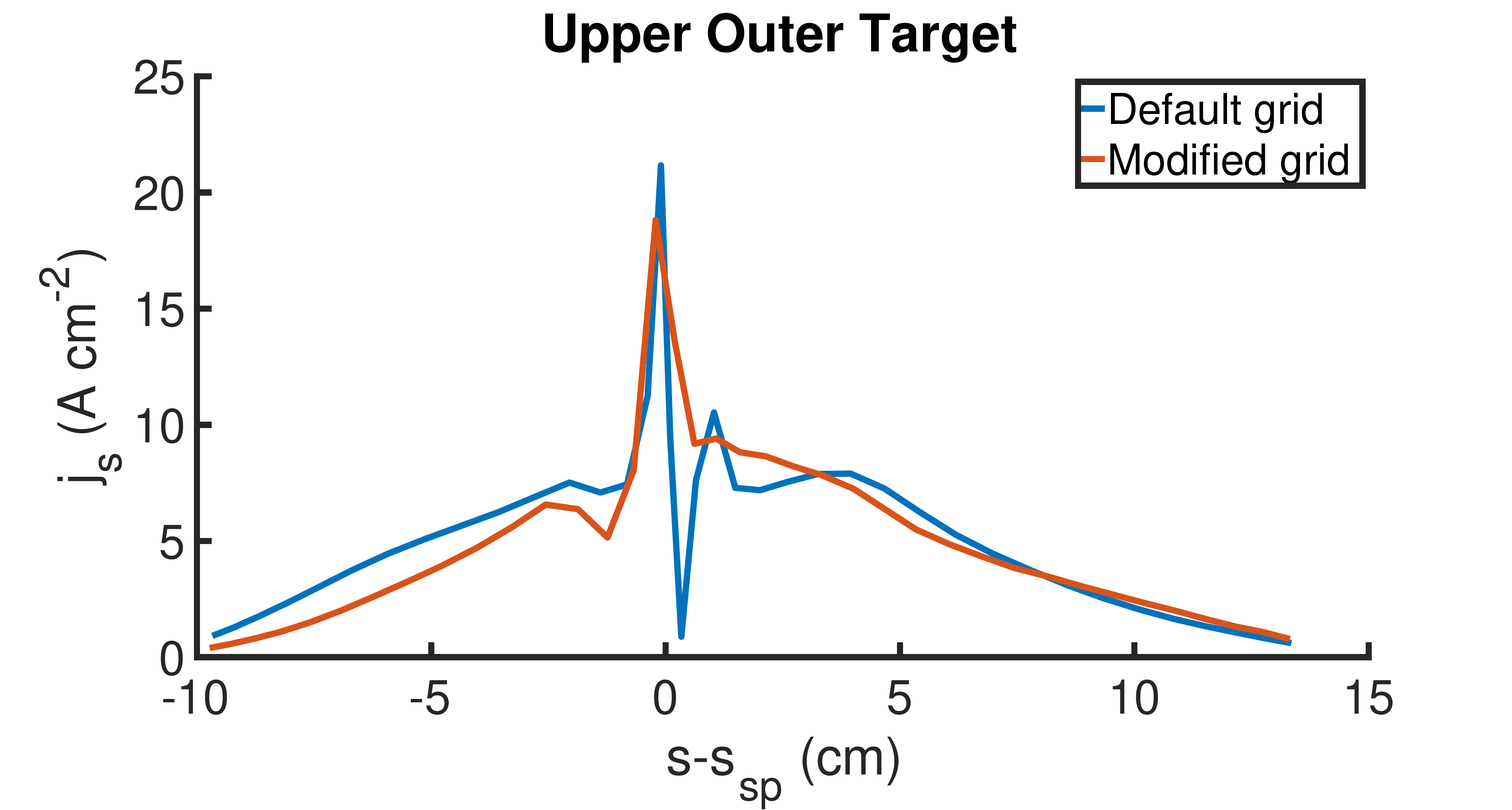}
	\caption{The influence of the grid on the ion saturation current at the UOT.}
	\label{fig:UOT_js_grid}
\end{figure}
\begin{figure}[h]
	\centering
	\begin{subfigure}{5.8cm}
		\includegraphics[width=5.8cm]{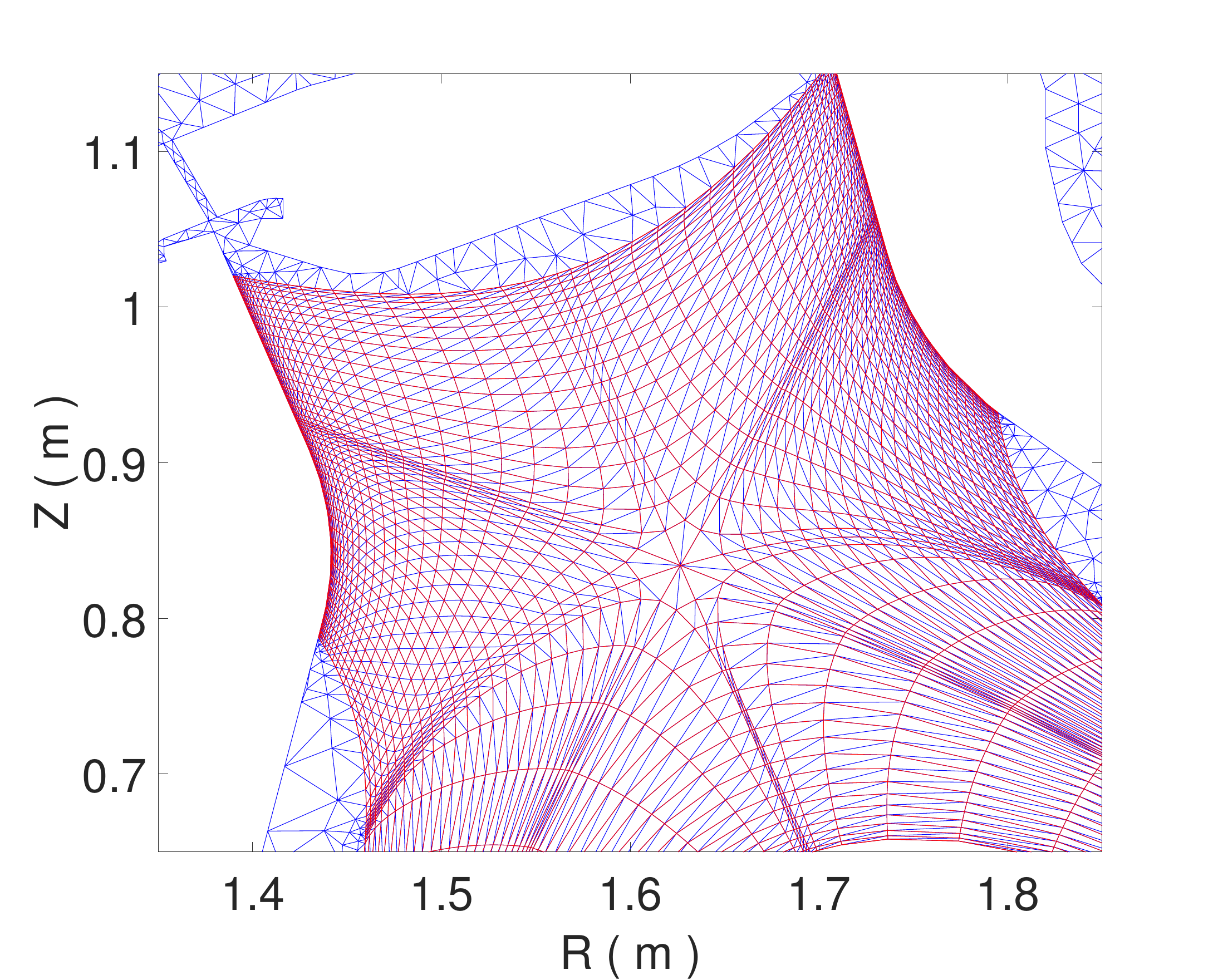}
		\caption{}
		\label{subfig:grid_original35_detail}
	\end{subfigure}
	\begin{subfigure}{5.8cm}
		\includegraphics[width=5.8cm]{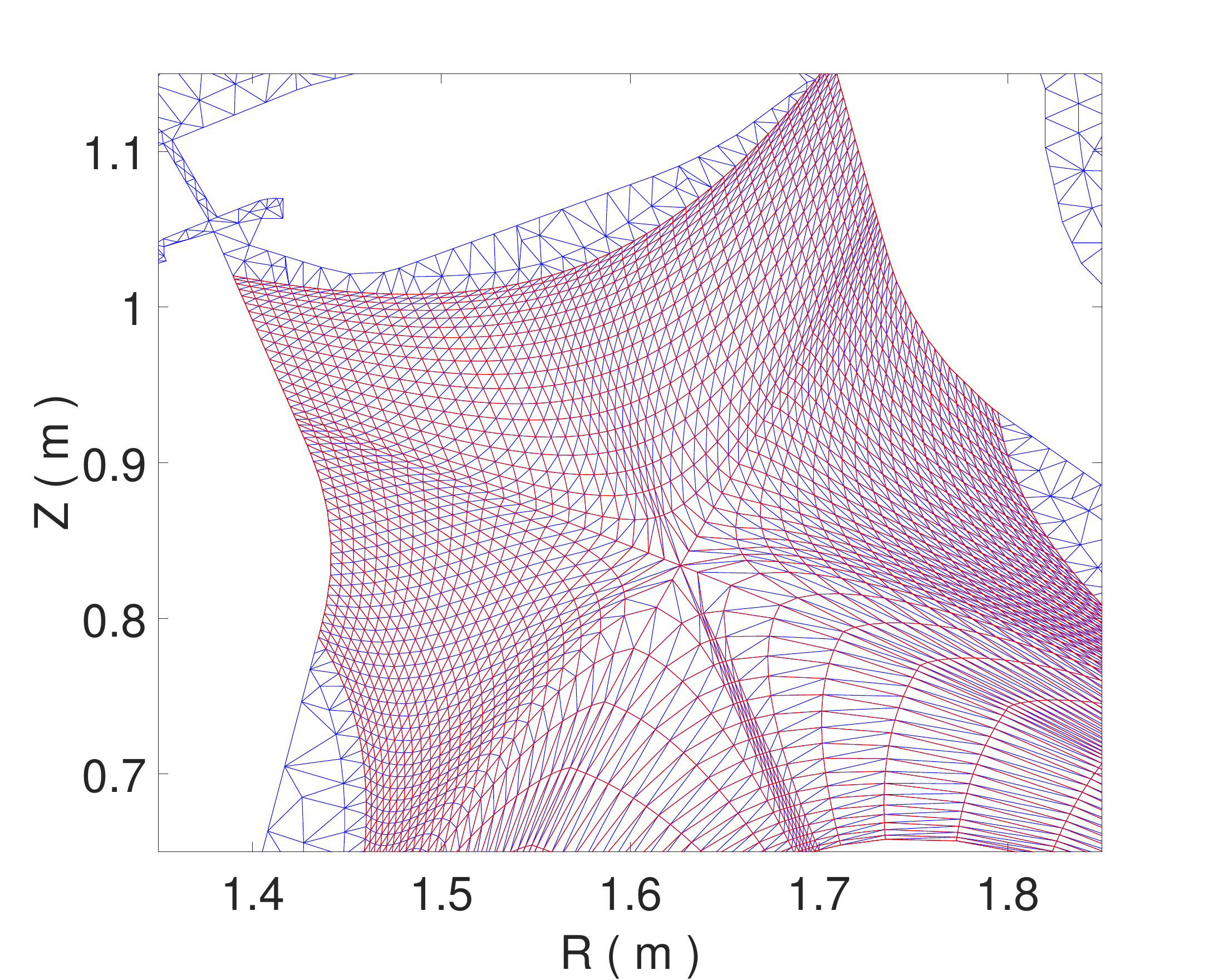}
		\caption{}
		\label{subfig:grid_changed35_detail}
	\end{subfigure}
	\caption{In (a) the default grid is shown in the vicinity of the X-point. In (b) the modified grid is shown at the same location. Both grids have the same amount of grid cells, but in the modified one, no refinement is done towards the divertor targets. The rectangular B2.5 grid is shown in red, and the triangular EIRENE grid in blue.}
	\label{fig:grid_drifts_detail}
\end{figure}

Figure \ref{fig:UOT_potential_grid} shows the potential at the UOT and figure \ref{fig:UOT_js_grid} the ion saturation current at the UOT in case of the default grid of ref. \cite{boeyaert2022numerical} (default grid) and in case of a grid with the same number of grid cells, but with a poloidal and radial grid width of minimum $1 \, \mathrm{mm}$ (modified grid).  Normally, SOLPS-ITER grids are refined towards the divertor targets as the plasma-wall interaction has a major influence on the simulation result. In the modified grid, this refinement is not performed as can be seen in figure \ref{fig:grid_drifts_detail}. In that way, grid cell widths smaller than $1 \, \mathrm{mm}$ are avoided. The modified grid, is the one used in all presented studies in this paper. As can be seen from figure \ref{fig:grid_drifts_detail}, both grids cover the same region in EAST. 

\begin{figure}[h]
	\centering
	\includegraphics[width=7cm]{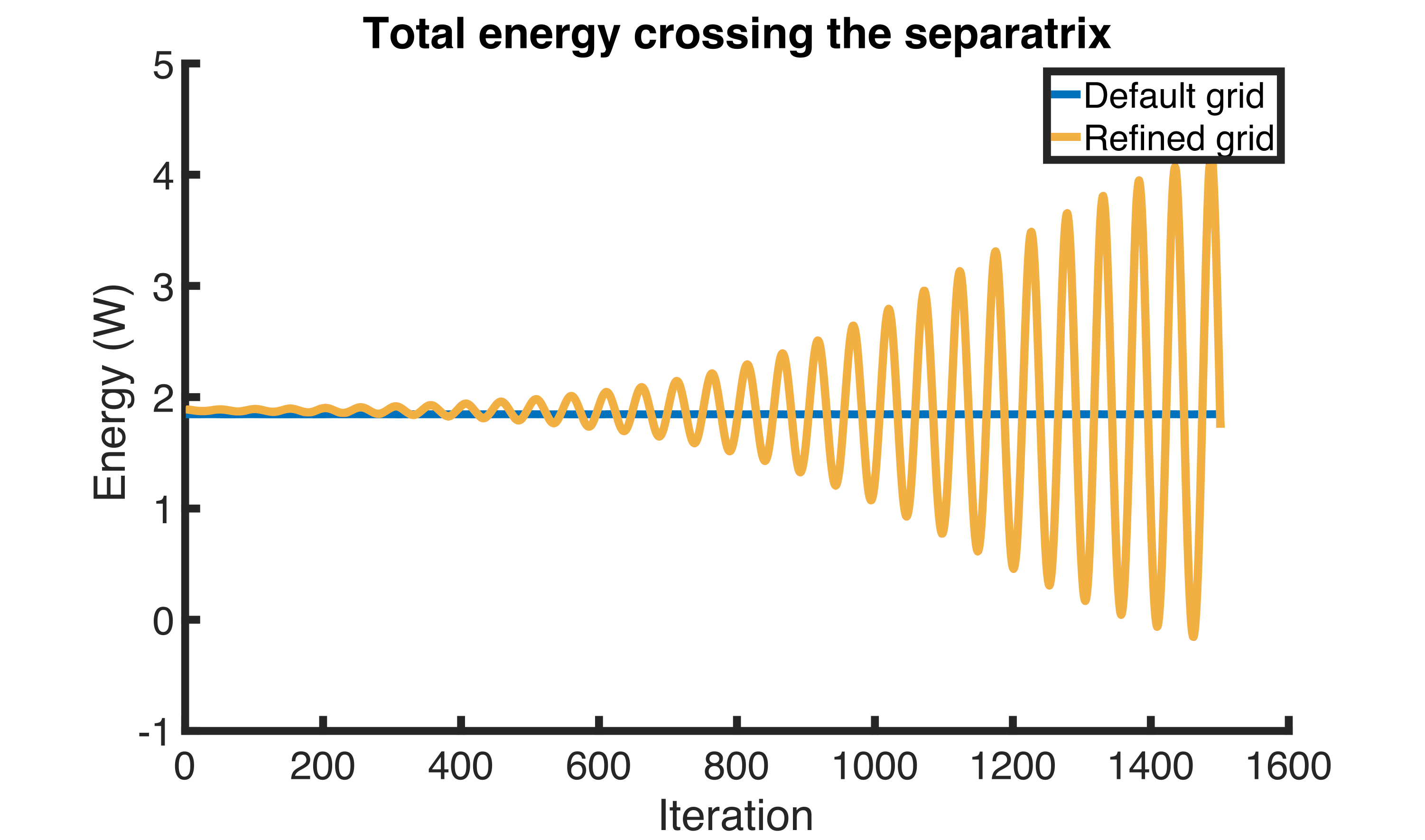}
	\caption{The appearing instabilities when the viscous terms are included. The instabilities on the energy crossing the separatrix for different grids are shown.}
	\label{fig:Energy_sep}
\end{figure}

While there are numerical oscillations in the potential calculation using the default grid, they are not present using the modified one. In the default grid, the minimal radial cell width is 1 mm and the poloidal one minimum 0.667 mm. 
In figure \ref{fig:Energy_sep} the energy crossing the separatrix is investigated for a further refined grid of ref. \cite{boeyaert2022numerical} (refined grid) in which also the radial cell width is below 1 mm. The figure shows that for such a refined grid, a numerical instability occurs in the simulation. 
This is in agreement with the findings of ref. \cite{wensing2021drift} where a similar instability appeared at the divertor targets for TCV simulations with radial grid cell widths below $1 \, \mathrm{mm}$.

Increasing further the minimal poloidal and radial grid width results in even smoother potential profiles. However, as this facilitates a larger discretization error, this is not investigated in detail. Coarsening the grid increases in that way the numerical diffusion which gives a similar effect as an increase $c_{\sigma_{\alpha,0}}$ and $c_{\alpha,0}$ discussed in the previous section. In the results presented in this paper, the modified grid with a radial and poloidal grid cell width of minimum 1 mm is employed.

\subsubsection{Influence of the number of included Monte Carlo particles and averaging}

\begin{figure}[h]
	\centering
	\includegraphics[width=7cm]{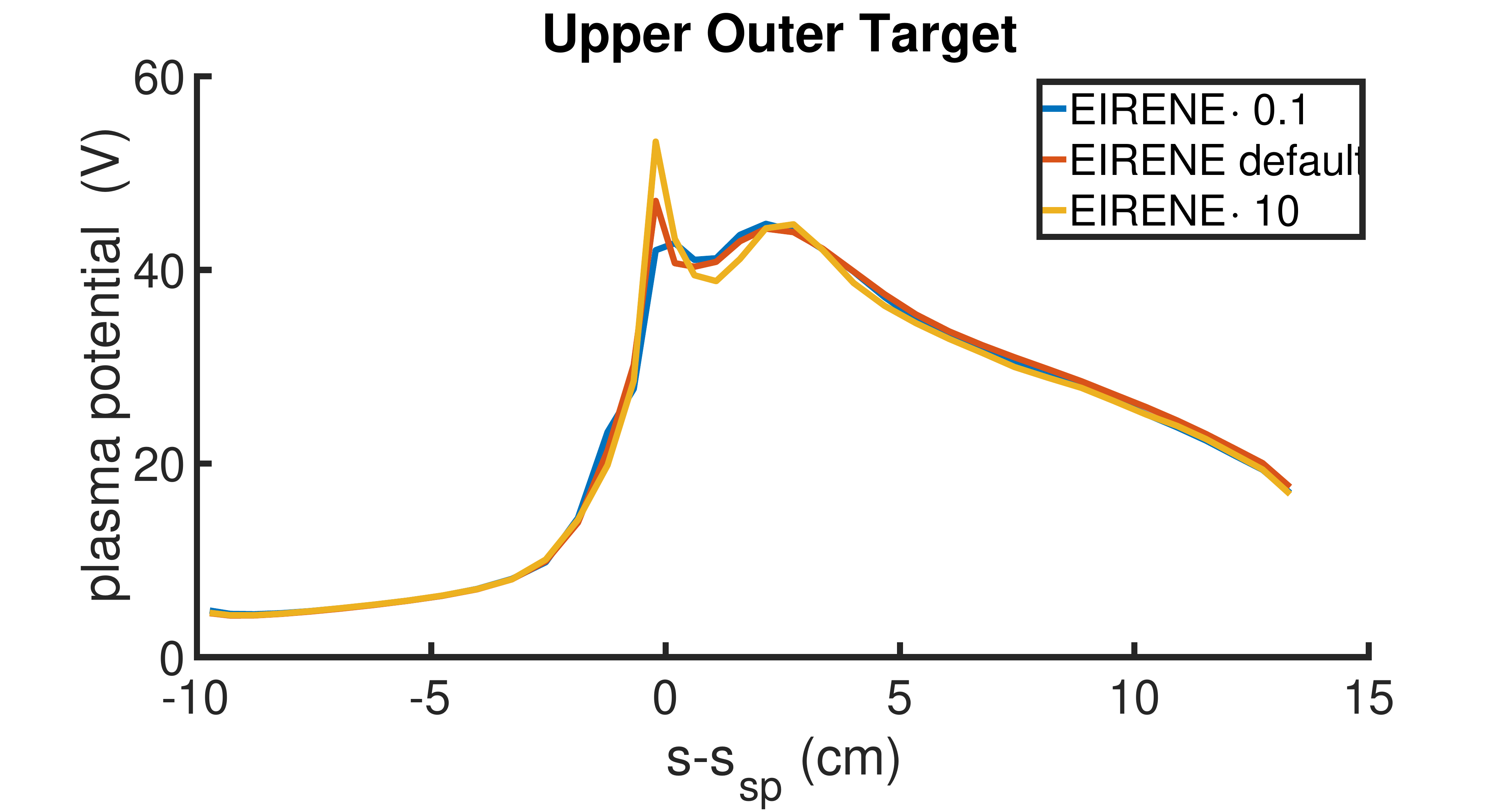}
	\caption{The influence of the number of involved EIRENE particles on the plasma potential at the UOT.}
	\label{fig:UOT_potential_EIRENE}
\end{figure}

Next, the influence of the number of involved Monte Carlo particles in the EIRENE simulations is investigated. The default number of EIRENE particles included in the presented simulations is based on table 1 of ref. \cite{boeyaert2022numerical} and the parameters used in the presented study are copied in table \ref{tab:EIRENE_particles}. After convergence, a simulation is averaged over 45,000 iterations to decrease the statistical error. In ref. \cite{boeyaert2022numerical} it is shown that this combination of the involved number of Monte Carlo particles and SOLPS-ITER averaging results in statistical errors below $0.35 \, \%$ on all investigated plasma quantities for a simulation without drifts. In figure \ref{fig:UOT_potential_EIRENE} the effect of including 10 times more or 10 times less Monte Carlo particles in the simulation with fully activated drifts is demonstrated. This indicates that the reduced fluctuations between different iterations -- resulting from an increased number of Monte Carlo particles --  influence the potential profile. 

\begin{table}[h!]
	\begin{tabular}{c| c}
		Place in the EIRENE grid & Number of EIRENE particles \\
		\hline
		Deuterium puff at OMP & 500\\
		Volumetric recombination & 700\\
		Recycling at the target surfaces & 500 (for each target)\\
		Recycling at the other walls of the B2.5 grid & 100 (for each wall)\\
	\end{tabular}
	\caption{The number of EIRENE particles per iteration involved in the simulations.}
	\label{tab:EIRENE_particles}
\end{table}

Figure \ref{fig:UOT_potential_EIRENE} supports the findings of the previous sections, namely that a higher potential drop around the separatrix is more realistic. It also indicates that with the default number of EIRENE particles, the potential drop around the separatrix is still underestimated. It should be noted that the analysis of ref. \cite{boeyaert2022numerical} showed that for a simulation without drifts, this number of EIRENE particles is sufficient.

\begin{figure}[h]
	\centering
	\includegraphics[width=7cm]{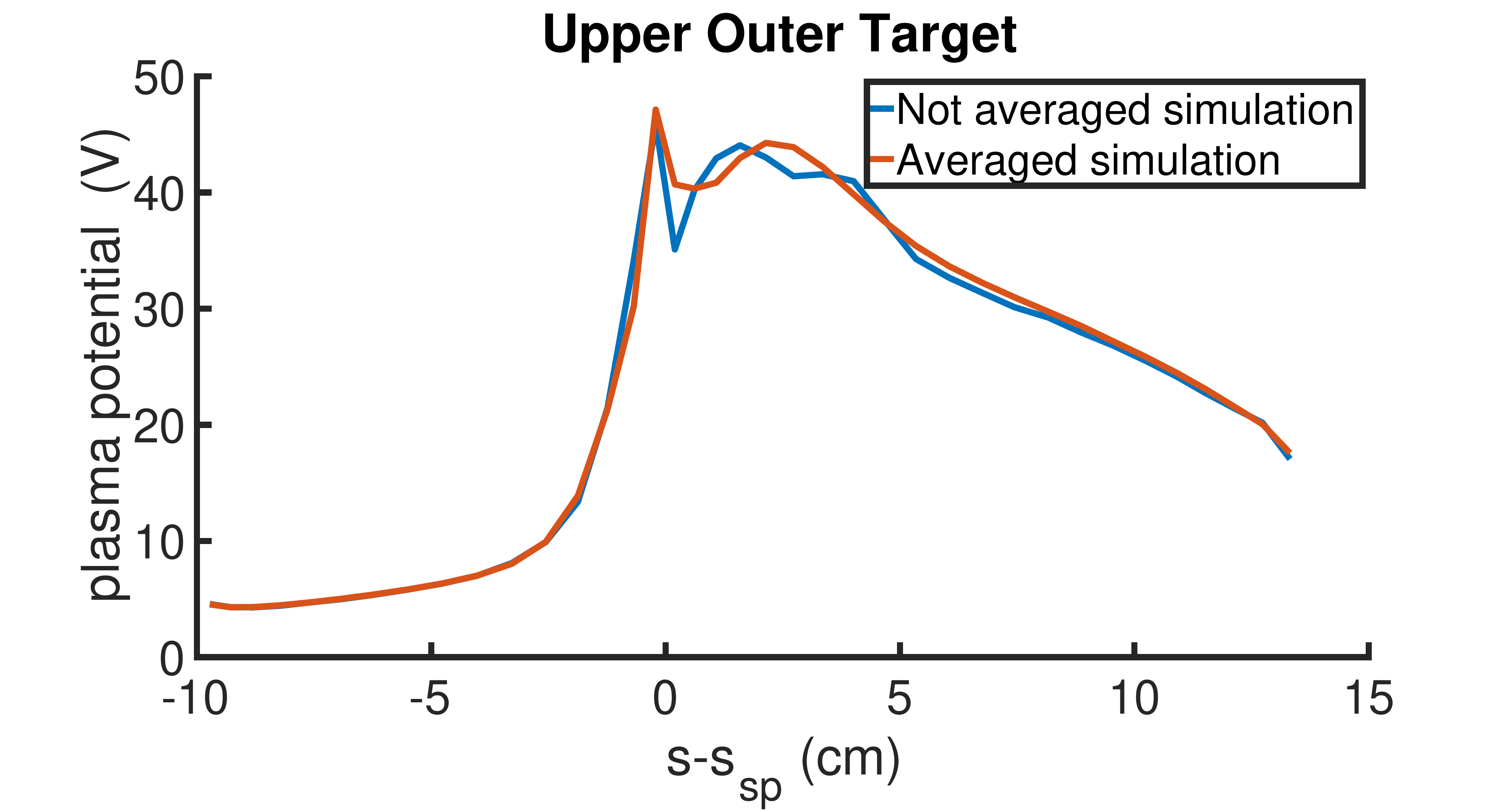}
	\caption{The influence of averaging on the plasma potential at the UOT.}
	\label{fig:UOT_potential_averaging}
\end{figure}

Figure \ref{fig:UOT_potential_averaging} shows the importance of SOLPS-ITER averaging over several simulations \cite{ghoos2020accuracy}. The figure shows that a larger potential drop is found for the non averaged simulations, but this is caused by a deeper drop around the separatrix while figure \ref{fig:UOT_potential_EIRENE} and the analysis of the previous section show that this is not expected (the potential should go higher before the separatrix, not lower afterwards). On top, also the profiles at other locations are not as smooth as in the averaged case.

\subsubsection{Employed time step and used numerical parameters to obtain convergence}
\label{subsubsec:employed_numerics}

\begin{figure}[h]
	\centering
	\begin{subfigure}{7cm}
		\includegraphics[width=7cm]{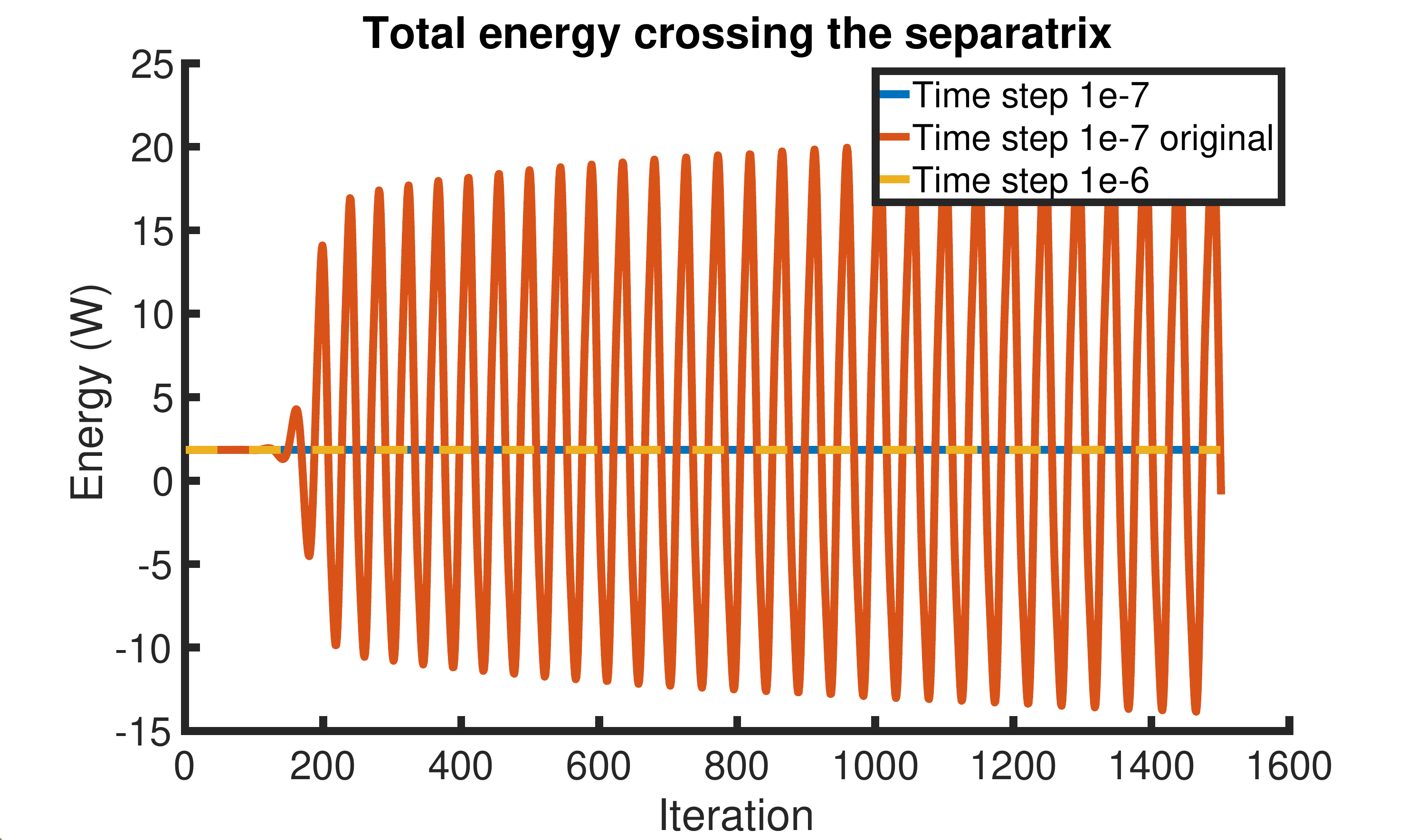}
		\caption{}
		\label{subfig:drift_instability_sep}
	\end{subfigure}
	\begin{subfigure}{7cm}
		\includegraphics[width=7cm]{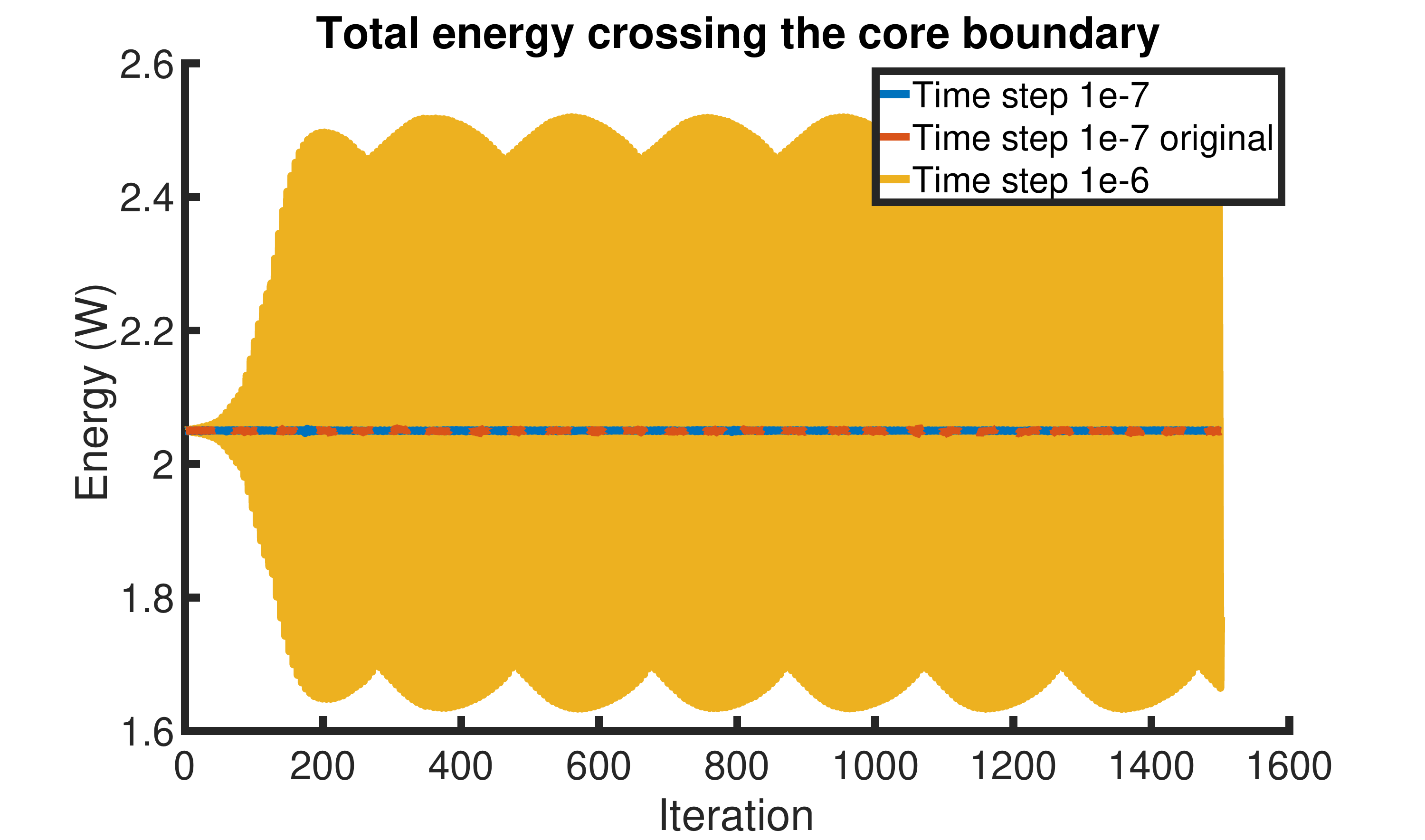}
		\caption{}
		\label{subfig:drift_instability_core}
	\end{subfigure}
	\caption{The instabilities appearing when the viscous terms are included. In (a) the instabilities on the energy crossing the separatrix for different setups are shown and in (b) the ones while crossing the core boundary of the grid. 'Original' in the legend refers to the default values for the boundary under-relaxation factors in SOLPS-ITER ($bc_{ref,te}$ = $bc_{ref,ti}$ = $0.01$). For the blue curves, these values are lowered to $bc_{ref,te}$ = $bc_{ref,ti}$ = $0.001$.}
	\label{fig:drift_instability}
\end{figure}

The next topic discussed is the required time step for the simulations. Where the previous sections showed influences on the final simulation result, the time step rather impacts the convergence itself. For the simulations without drifts from ref. \cite{boeyaert2022numerical}, a time step of maximum $5.0 \cdot 10^{-5} \, \mathrm{s}$ is required to obtain a stable solution. However, this is not possible for simulations in which the viscous terms are activated. In figure \ref{fig:drift_instability}, the variations in time of the total energy crossing the separatrix (a) and core boundary (b) are given for simulations in which the viscous terms are activated. As can be seen in yellow on figure \ref{subfig:drift_instability_core}, with a time step of $1.0 \cdot 10^{-6} \, \mathrm{s}$ the simulation does not converge and gives numerical oscillations for the energy crossing the core grid boundary. A time step of maximum $1.0 \cdot 10^{-7} \, \mathrm{s}$ is needed (the blue line in both plots of figure \ref{fig:drift_instability}). The requirement of a small time step can be explained by the segregation of all different equations and the intrinsic noisy interaction with the EIRENE Monte Carlo code. This requires small time steps to avoid numerical instabilities in drift cases as the maximum time step is related to the convergence of the process with the smallest time scale involved in the numerical problem \cite{wensing2021drift}. In SOLPS-ITER this means that a higher time step leads to divergence in the solution of the potential equation \cite{kaveeva2018speed}. 

In ref. \cite{kaveeva2018speed} the appearing numerical instabilities are explained by "the poloidal redistribution of particles inside the separatrix by $\textbf{E}$ x $\textbf{B}$ drifts in combination with the modification of the radial electric field by diamagnetic currents". These observations are obtained for ASDEX Upgrade and ITER simulations. As indicated in figure \ref{fig:drift_instability}, a similar instability happens in the studied EAST discharges. The energy balance is examined during all simulations indicating that the energy crossing the core boundary of the grid does not stay stable. For the performed drift simulations, the energy crossing the core boundary is imposed as a boundary condition for the ion and electron energy equations. As such a flux BC is from the numerical viewpoint less stable than a Dirichlet type BC, drift simulations in SOLPS-ITER  require a type of BC in which the code converts these imposed heat fluxes to electron and ion temperatures \cite{bonnin2020solps}. During a series of iterations, the new imposed temperature will depend on the actual fluxes imposed at the BC, the ones at the previous time step, and on how the imposed temperature should be corrected to match the heat flux from the BC:

\begin{equation}
	\begin{split}
		T_{e,new} = T_{e,old} \cdot \left(1 - bc_{ref,te} \cdot (fh_e - f_{imposed,e}) \right) \\
		T_{i,new} = T_{i,old} \cdot \left(1 - bc_{ref,ti} \cdot (fh_i - f_{imposed,i}) \right).
	\end{split}
\end{equation}

In this process, the importance of the previous time step can be regulated with under-relaxation factors -- $bc_{ref,te}$ and $bc_{ref,ti}$, called boundary under-relaxation factors in the further scope of this paper. $fh_e$ and $fh_i$ are the actual power fluxes going through the core boundary, where $f_{imposed,e}$ and $f_{imposed,i}$ are the ones imposed in the boundary condition. This effect is demonstrated with the red curve in figure \ref{fig:drift_instability}: if the influence of the previous time step is too large, the simulation will not converge. The fact that the influence of the previous time step is limited, also contributes to the long converging times required for drift simulations.

Additionally to the imposed under-relaxation factors, also an appropriate choice of the numerical parameters is needed. In a simulation in which no drifts are present, the plasma equations are solved with a time step 100 times higher inside the separatrix for the ion and electron energy equations, and a time step 10 times higher for the continuity equation. As the critical point in drift simulations lies inside the separatrix, this speed-up method does not lead to a converged simulation anymore. Therefore, the time step inside the separatrix should be lower than the one outside the separatrix as discussed in ref. \cite{kaveeva2018speed}.



\subsection{Including drifts in a Ne-seeded attached simulation}


In a next step, Ne-seeding is added to SOLPS-ITER simulations. As neon will also radiate in the core, the input power crossing the core boundary of the SOLPS-ITER grid is decreased from $2.05 \, \mathbf{MW}$ to $1.80 \, \mathbf{MW}$. Other boundary conditions are kept the same as in the unseeded simulation. If only a small amount of Ne is added, the simulations are still attached. This makes it possible to examine the influence of the presence of impurities on the numerics in the SOLPS-ITER code. More precisely the time step and numerical parameters influencing the convergence are examined in the presented work.

\begin{figure}[h]
	\centering
	\includegraphics[width=7cm]{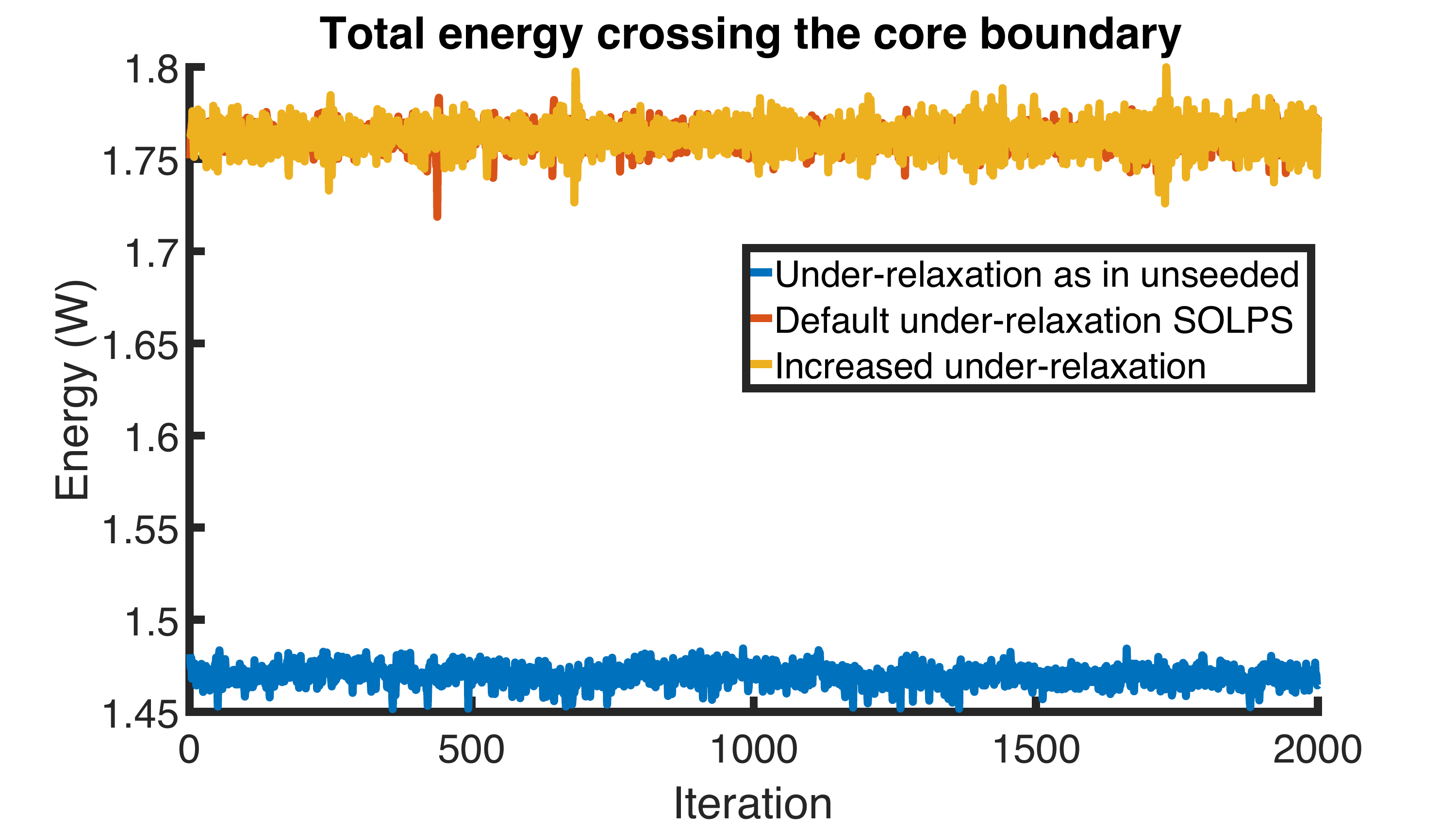}
	\caption{The effect on the energy crossing the core boundary caused by the employed under-relaxation factors in the boundary conditions.}
	\label{fig:drift_instability_Ne}
\end{figure}

\begin{figure}[h]
	\centering
	\includegraphics[width=7cm]{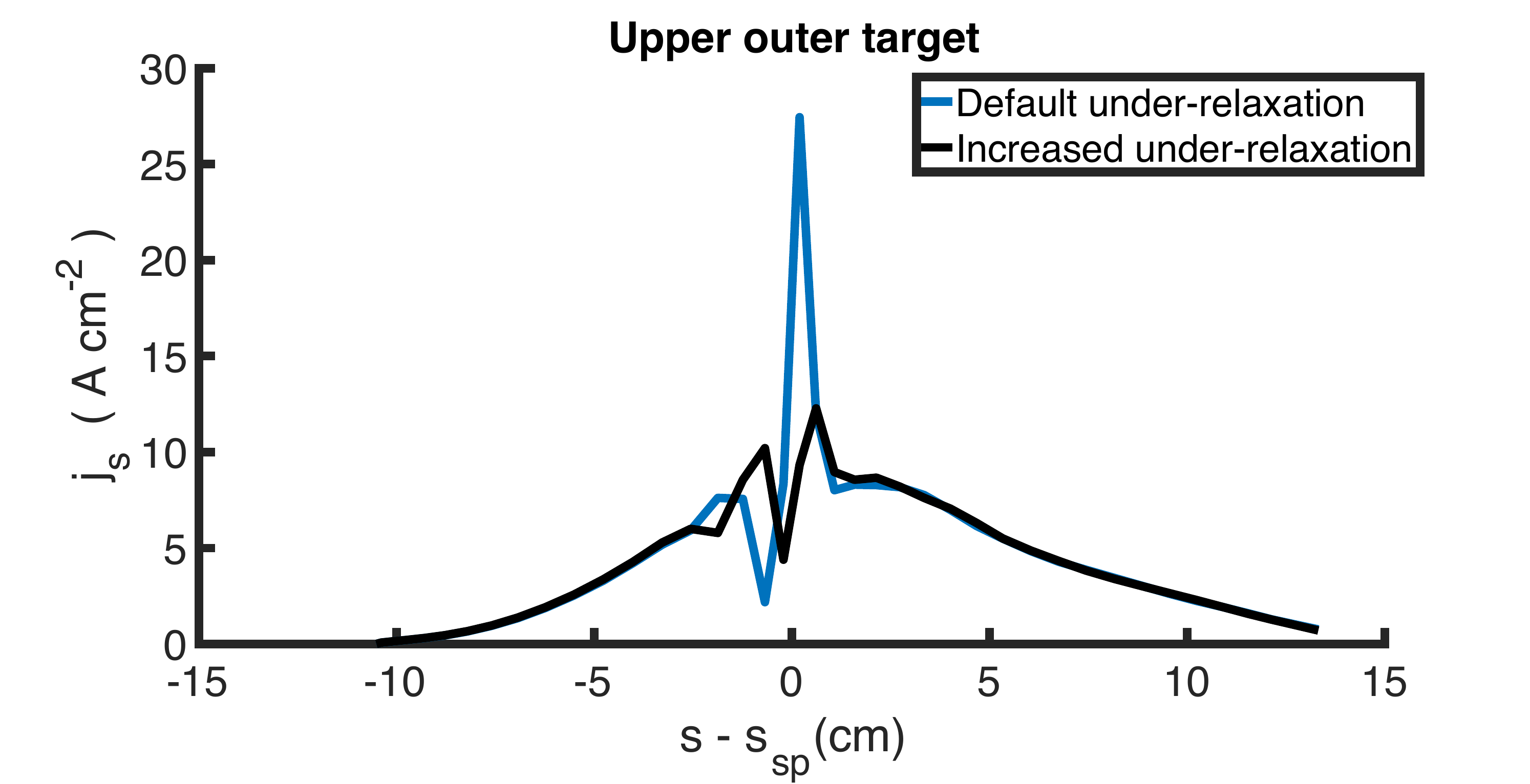}
	\caption{Influence of the under-relaxation factor on $j_s$ at the UOT in attached Ne-seeded simulations}
	\label{fig:DivLP_SOLPS_js_UOT_b2stbc_bc_ref_tx_js_Ne}
\end{figure}

In figure \ref{fig:drift_instability} it was demonstrated that a decreased boundary under-relaxation factor is required to converge the simulations. Figure \ref{fig:drift_instability_Ne} shows the opposite effect if Ne-seeding is present: a too low under-relaxation factor decreases the influence of the current time step too much. This makes that the simulation does not converge to the correct energy crossing the core boundary of the employed grid. In case the under-relaxation factor is increased, the power crossing the core boundary increases up to $98 \, \%$ of the power which is imposed at the core boundary. If the boundary under-relaxation factor is increased too much, however, there is a discrepancy between the imposed power and the result of the simulation. Figure \ref{fig:drift_instability_Ne} shows that a boundary under-relaxation factor of 0.03 still gives a correct power crossing the core boundary.  In figure \ref{fig:DivLP_SOLPS_js_UOT_b2stbc_bc_ref_tx_js_Ne} the effect of the boundary under-relaxation factor on the ion saturation current at the UOT is shown for the Ne-seeded simulation. This indicates that a larger boundary under-relaxation factor leads to lower peaks as the set of equations is numerically more stable. Therefore, a boundary under-relaxation factor of 0.03 is used for the performed Ne-seeded simulations.

It should be remarked that these simulations were performed with 10 times less EIRENE particles than the unseeded simulations presented earlier in this paper. Figure \ref{fig:UOT_potential_EIRENE} shows that the potential in an unseeded simulation is smoother if less EIRENE particles are used. 



\subsection{Including drifts in a Ne-seeded detached simulation}

\begin{figure}[h]
	\centering
	\includegraphics[width=7cm]{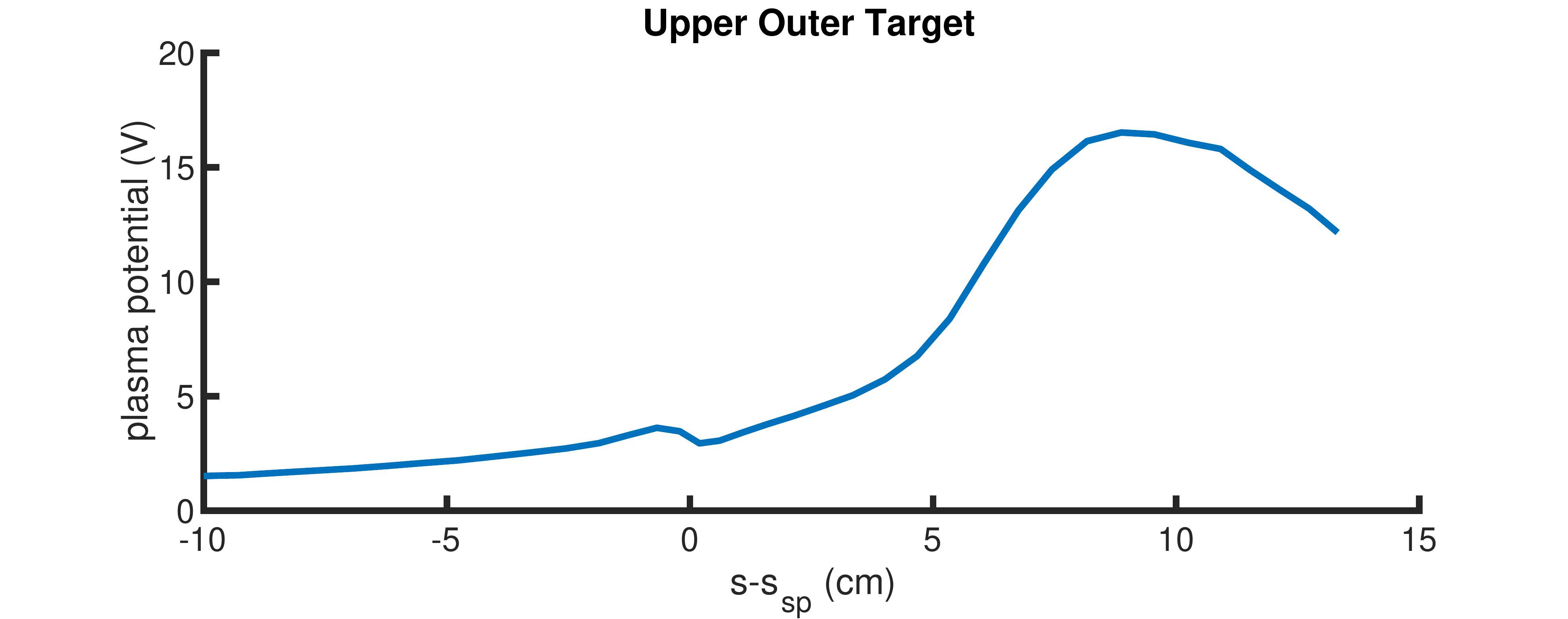}
	\caption{The plasma potential at the UOT in detachment.}
	\label{fig:UOT_potential_detached}
\end{figure}

Figure \ref{fig:plasma_potential} showed a decreased potential in detached simulations. Nevertheless, the potential is not completely smooth as demonstrated in figure \ref{fig:UOT_potential_detached} with the potential profile at the UOT in a simulation in which enough Ne is added to bring the plasma in detachment. Also for this simulation 10 times less EIRENE particles are used due to the limited available computing time. In contrast to the earlier shown simulations, temperature boundary conditions are applied at the innermost grid boundary. This means that no under-relaxation factors are needed. However, this makes it more difficult to mimic the physics of a real experiment as the input power is known more precisely than the temperature at the core boundary of the B2.5 grid.


%

\section{Effect of drifts on the power crossing the grid boundary towards the main wall}

\begin{figure}[h]
	\centering
	\includegraphics[height=8cm]{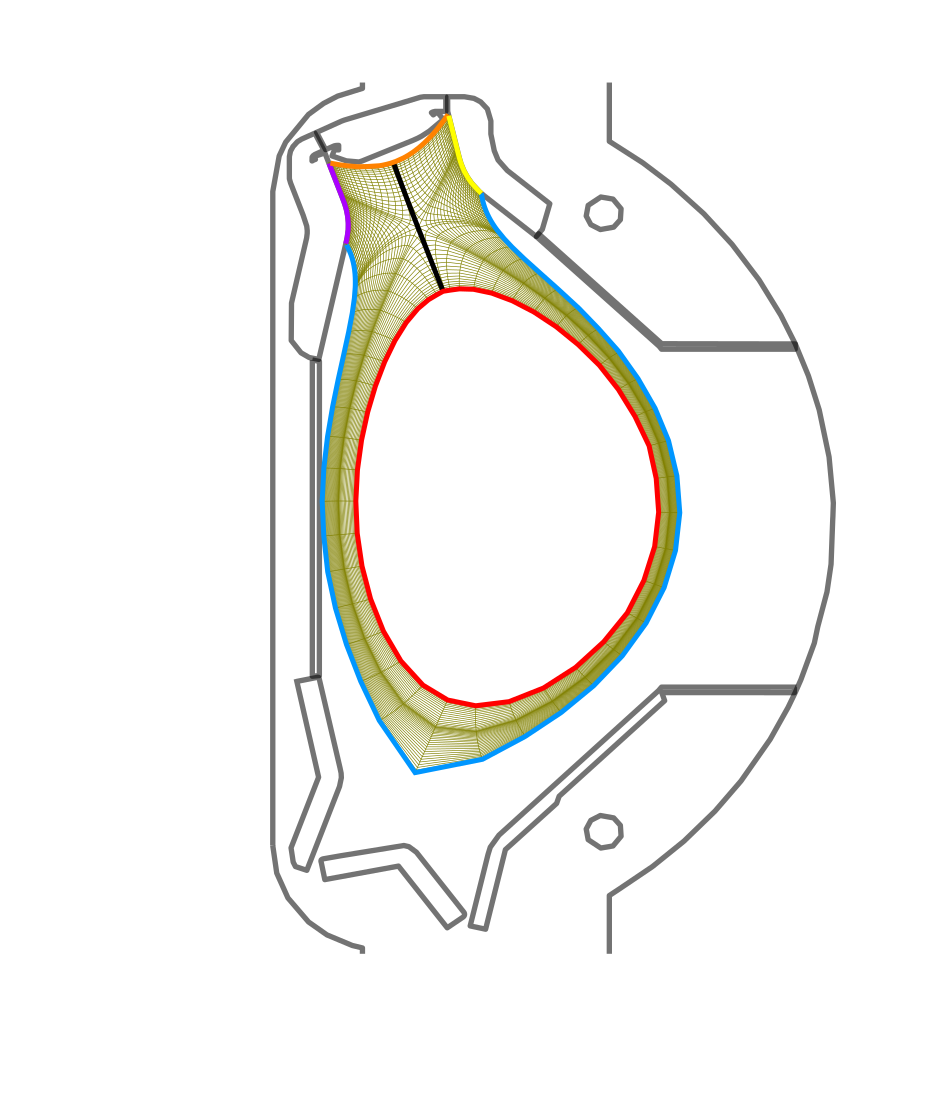}
	\caption{Physical mesh for the plasma in the performed SOLPS-ITER simulations. The grid boundary towards the main chamber wall is indicated in blue. Through this boundary, the energy leaving the simulation domain should be limited to draw physical conslusions from the simulation.}
	\label{fig:boundary_conditions}
\end{figure}

As the grids in the 3.0.7 master version of SOLPS-ITER are not extended up to the first wall, it should be verified that they cover the region of the tokamak edge where the most significant fraction of physical processes take place. Ref. \cite{uljanovs2017isotope} showed for JET that, if the grid of the simulation is too narrow, it is difficult to draw physical conclusions  as the decay-length type BCs impact the solution too much and the plasma state is determined by these BCs. In order to determine if the grid extends far enough in the SOL, it is important to investigate how the injected power is dissipated. If too much power is deposited on the B2.5 grid boundary towards the main chamber wall, the simulation will not be able to give an accurate description of the power deposition in the divertor. Therefore, the power balance is verified. Special attention is put on the effect of drifts on the power crossing the grid boundary closest to the first wall as indicated in figure \ref{fig:boundary_conditions}. For the examined simulations (with and without drifts) the power crossing this grid boundary is summarized in table \ref{tab:energy_to_main_champber}. As it came out that the number of EIRENE particles does not influence the power balance, all percentages summarized in table \ref{tab:energy_to_main_champber} are given for simulations with 10 times less particles as in that case simulations for all examined conditions were available.

\begin{table}[h!]
	\centering
	\begin{tabular}{m{0.1\textwidth} c | c  }
		\hline
		& \multicolumn{2}{c}{Attached, purely deuterium simulations, energy BCs} \\
		& \multicolumn{2}{c}{input power = 2.05 MW}\\
		\hline
		& No drifts & $100 \, \%$ drifts\\
		Towards main chamber wall & 0.48 MW ($23.4 \, \%$) & 0.46 MW ($22.4 \, \%$)\\
		Towards outer divertor &  0.76 MW ($37.2 \, \%$) &  0.90 MW ($43.8 \, \%$) \\
		Towards inner divertor & 0.46 MW ($22.6 \, \%$) &  0.16 MW ($7.9 \, \%$)\\
		\hline
		\hline
		&  \multicolumn{2}{c}{Attached simulations with Ne-seeding, energy BCs}\\
		& \multicolumn{2}{c}{input power = 1.8 MW}\\
		\hline
		& No drifts & $100 \, \%$ drifts \\
		Towards main chamber wall & 0.47 MW ($26.1 \, \%$) & 0.44 MW ($24.4 \, \%$)\\
		Towards outer divertor & 0.63 MW ($34.8 \, \%$) &  0.74 MW ($40.9 \, \%$)\\
		Towards inner divertor & 0.44 MW ($24.1 \, \%$)& 0.13 MW ($7.21 \, \%$)\\
		\hline
		\hline
		& \multicolumn{2}{c}{Detached simulations with Ne-seeding, $T_e$ BCs}\\
		\hline
		& \multicolumn{2}{c}{$100 \, \%$ drifts} \\
		Towards main chamber wall & \multicolumn{2}{c}{0.35 MW ($19.5 \, \%$)} \\
		Towards outer divertor & \multicolumn{2}{c}{0.43 MW ($23.7 \, \%$)} \\
		Towards inner divertor & \multicolumn{2}{c}{0.20 MW ($11.0 \, \%$)} \\
		\hline
	\end{tabular}
	\caption{Percentage of energy crossing towards the main chamber wall towards the total energy entering the SOL.}
	\label{tab:energy_to_main_champber}
\end{table}

\begin{figure*}
	\begin{subfigure}{6cm}
		\includegraphics[width=6cm]{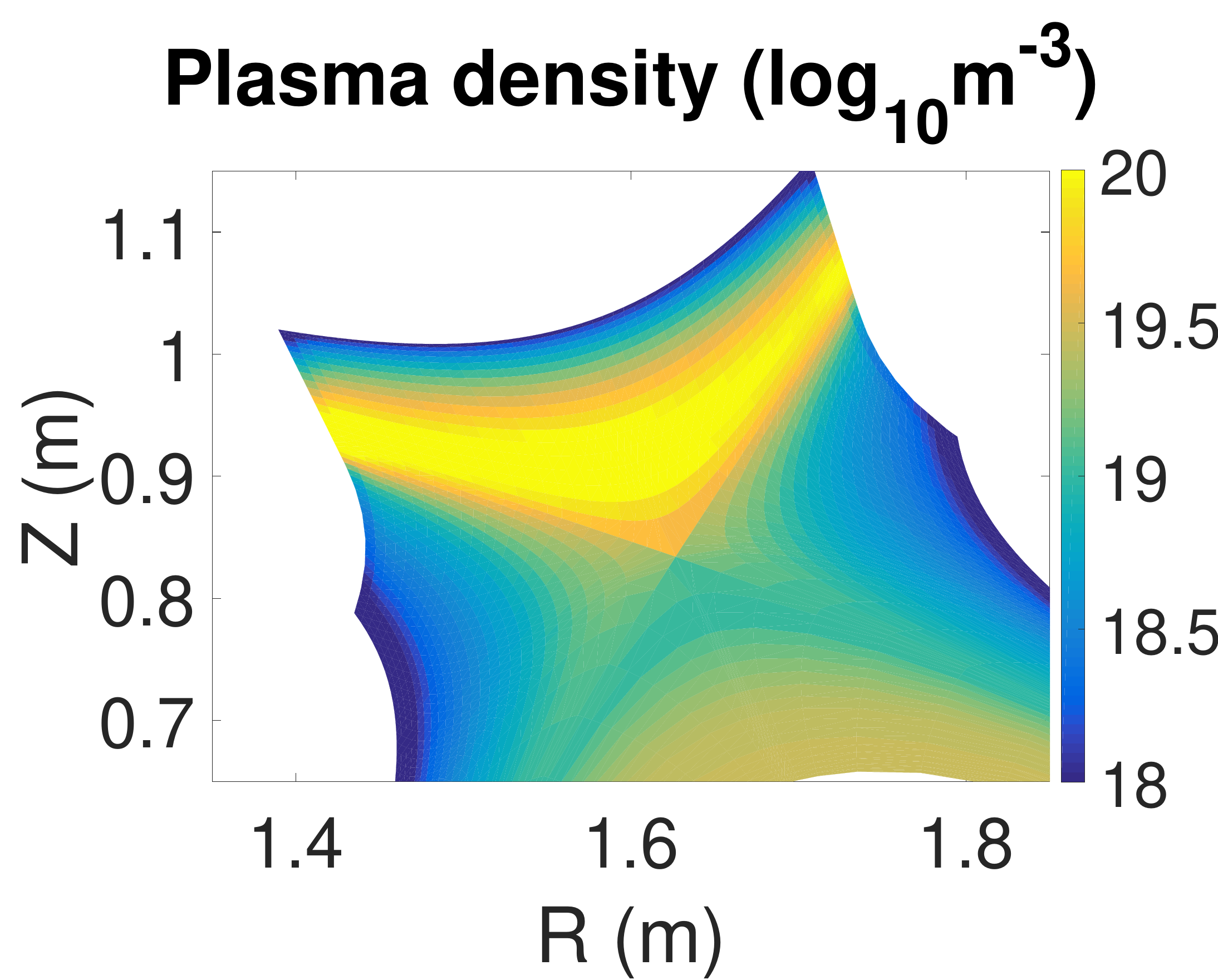}
		\caption{2D profile of $n_e$ without drifts}
		\label{subfig:SOLPS_2D_ne_nodrifts}
	\end{subfigure}
	\begin{subfigure}{6cm}
		\includegraphics[width=6cm]{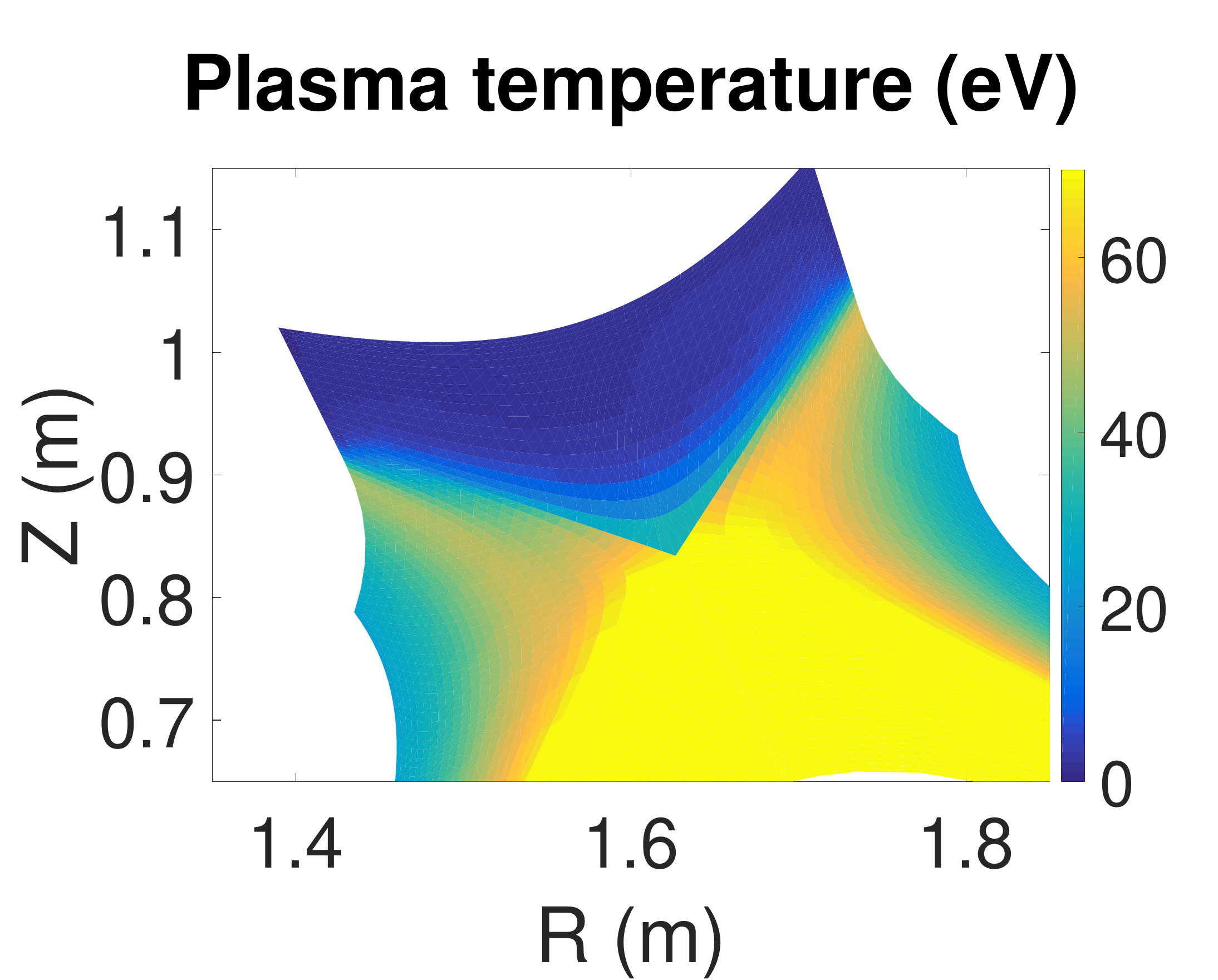}
		\caption{2D profile of $T_e$ without drifts}
		\label{subfig:SOLPS_2D_Te_nodrifts}
	\end{subfigure}
	\begin{subfigure}{6cm}
		\includegraphics[width=6cm]{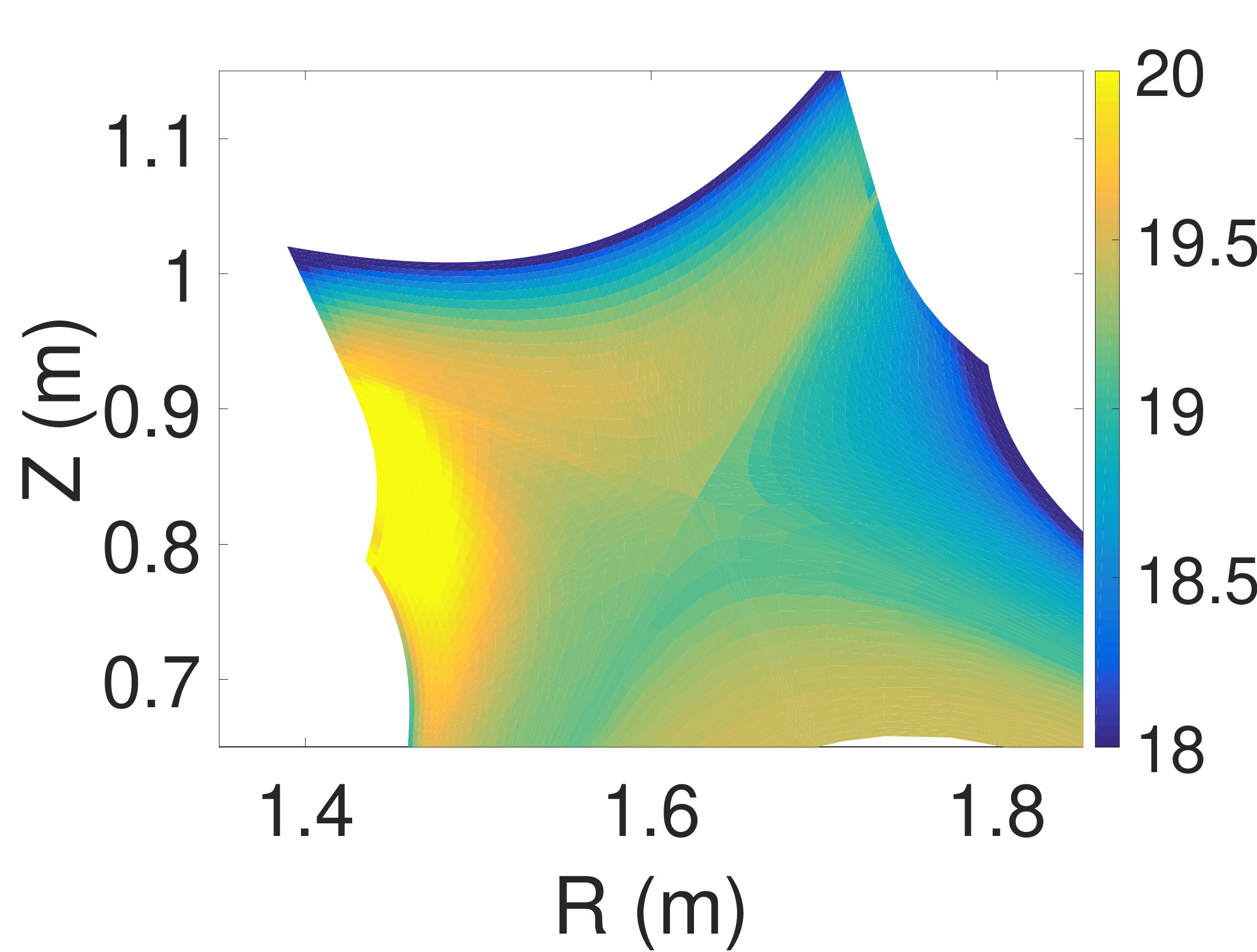}
		\caption{2D profile of $n_e$ with $100 \, \%$ drifts}
		\label{subfig:SOLPS_2D_ne_drifts}
	\end{subfigure}
	\begin{subfigure}{6cm}
		\includegraphics[width=6cm]{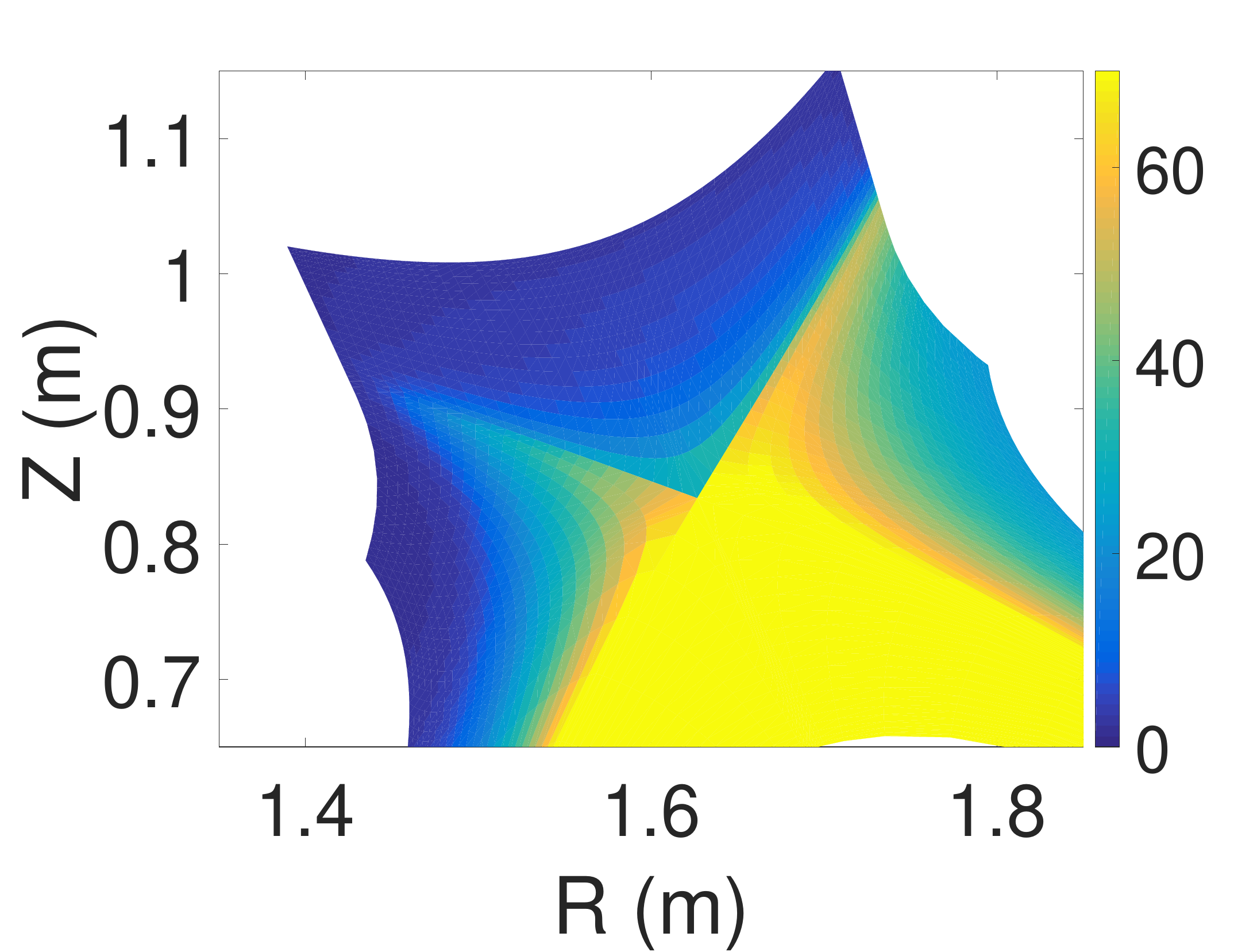}
		\caption{2D profile of $T_e$ with $100 \, \%$ drifts}
		\label{subfig:SOLPS_2D_Te_drifts}
	\end{subfigure}
	\caption{The 2D profiles in the vicinity of the upper divertor for $n_e$ and $T_e$ in the final simulation without (a,b) and with (c,d) drifts.}
	\label{fig:SOLPS_2D_ne_and_Te}
\end{figure*}

This overview demonstrates that in all simulations $\sim 75 \, \%$ of the energy crossing the core boundary of the grid (in red in figure \ref{fig:boundary_conditions}), is not deposited on the first wall. 
In the deuterium simulations, most of the energy will be transported towards the inner and outer divertor legs. In the Ne-seeded simulations, on the other hand, a significant part of the power will already be radiated before the entry of the divertor. 
The power sharing between inner and outer divertor shown in table \ref{tab:energy_to_main_champber} clearly indicates that drifts are required to obtain an asymmetry between inner and outer divertor. In the presence of drifts, only a small fraction of the power is going towards the inner divertor, and most of the power enters the outer divertor leg. This is also nicely illustrated on the 2D plots of figure \ref{fig:SOLPS_2D_ne_and_Te}, which show the 2D profiles for the density and temperature with and without drift. In case drifts are included, a clear asymmetry between the inner and outer divertor is present while this is absent when drifts are not switched on. As indicated before, this is in agreement with findings on other devices \cite{wensing2020experimental, dekeyser2017solps}.

The remaining $\sim 25 \, \%$ of the energy transported from the core into the scrape-off layer will be dissipated volumetrically outside the B2.5 grid and eventually mainly deposited on the first wall, and is not considered further in the simulation. A simulation with a grid extended until the first wall as introduced in ref. \cite{dekeyser2021plasma} for SOLPS-ITER, would make it possible to investigate where this energy is exactly deposited. However, as the simulation domain for the current simulations covers $73.9 \, \%$ to $80.5 \, \%$ of the power deposition, they can all (both without and with drifts) be used to investigate more in detail the power deposition in the divertor of EAST.

\section{Summary}

In this paper the numerical implications of including drifts in the SOLPS-ITER simulations of EAST are studied. For that, three simulations in upper single null configuration are analyzed: an attached purely deuterium simulation, an attached simulation in which low levels of Ne-seeding are present, and a detached Ne-seeded simulation. As the gradients in potential profiles strongly influence the plasma flows, and as these potential gradients are larger during attachment, the attached drift simulations are more difficult to converge than the detached ones. Therefore, also the influence of employed numerical parameters in the SOLPS-ITER setup are larger.

In order to converge SOLPS-ITER drift simulations, artificial anomalous terms are introduced in the charge continuity equation. These terms are determined by an artificial anomalous conductivity and an artificial anomalous thermo-electric coefficient. In the presented work it is shown that the choice of these parameters largely influence the potential, and in that way the E x B plasma flows. The diamagnetic drifts, on the other hand, are independent of these artificial anomalous parameters.

In a next section, the influence of the grid on the plasma potential is verified. It came out that grid cell widths below 1 mm give raise to numerical instabilities which are reflected in the potential profile. Therefore, it is recommended to use grid cell widths of at least 1 mm in the radial and poloidal direction.

The number of EIRENE particles in the SOLPS-ITER simulation also influences the potential profile. Around the separatrix an abrupt change in the plasma potential is expected. However, this peak can give numerical issues as the Monte Carlo noise of the EIRENE simulation resulting from too few particles gives rise to local fluctuations which may impact convergence. Therefore, a larger number of EIRENE particles is necessary. 
This has as a drawback that the required computing time increases. Additionally, applying SOLPS-ITER averaging after convergence is reached, smooths the potential profiles.


Next, the time step and involved under-relaxation factors for the imposed energy boundary conditions which are needed to obtain a converged simulation are described.

After finishing the analysis for the attached purely deuterium simulation, the attached Ne-seeded and detached Ne-seeded simulations are analyzed. The choice of under-relaxation factors for the energy boundary condition seems to be dependent on the sepcific case: the attached Ne-seeded simulation only converged correctly if increased under-relaxation factors were imposed.

In a last section, the effect of drifts on the power crossing the grid boundary towards the main wall is investigated. As the employed SOLPS-ITER grid is not extended up to the first wall, it is important to keep track of this to avoid an artificial loss of power through the outermost boundary of the B2.5 grid. The performed analysis shows that the simulations can be used for divertor power deposition studies of the investigated EAST discharges.

\begin{acknowledgments}
The authors would like to thank the EAST team, in particular Yunfeng Liang, Liang Wang, Fang Ding and Kedong Li, for providing the EAST related parameters (equilibrium, geometry, pump efficiency...) used in the presented study.

This work has partially been carried out within the EU-CN collaboration framework of the EUROfusion Consortium, funded by the European Union via the Euratom Research and Training Programme (Grant Agreement No 101052200 — EUROfusion). Views and opinions expressed are however those of the author(s) only and do not necessarily reflect those of the European Union or the European Commission. Neither the European Union nor the European Commission can be held responsible for them.

This work was funded in parts by the U.S. Department of Energy (DE-SC0014210) and the College of Engineering at UW Madison, WI, USA

The SOLPS-ITER simulations were performed at the Marconi supercomputer from the National Supercomputing Consortium CINECA.
\end{acknowledgments}

\nocite{*}
\bibliography{papers_drifts}

\end{document}